\documentclass[smallextended]{svjour3}       
\smartqed  
\usepackage{graphicx}
\usepackage{amssymb,amsmath,graphicx,color,textcomp}
\usepackage{amsfonts}
\usepackage{subfig,caption}
\usepackage{epsfig,latexsym,graphicx}
\begin{document}

\title{Robust Estimation of Bivariate Tail Dependence Coefficient 
}


\author{Abhik Ghosh       
}


\institute{A. Ghosh \at
              Bayesian and Interdisciplinary Research Unit\\
              Indian Statistical Institute, Kolkata, India.\\
              \email{abhianik@gmail.com}           
}

\date{Received: date / Accepted: date}

\maketitle

\begin{abstract}
The problem of estimating the coefficient of bivariate tail dependence is considered here 
from the robustness point of view; it combines two apparently contradictory theories of 
robust statistics and extreme value statistics.
The usual maximum likelihood based or the moment type estimators of tail dependence coefficient are 
highly sensitive to the presence of outlying observations in  data. 
This paper proposes some alternative robust estimators
obtained by minimizing the density power divergence with suitable model assumptions; 
their robustness properties are examined through the classical influence function analysis.
The performance of the proposed  estimators is illustrated through an extensive empirical study 
considering several important bivariate extreme value distributions. 
\end{abstract}

\section{Introduction}

In case of several applied sciences including economics, finance, hydrology etc., 
modeling the rare events is becoming very important now-a-days; it is essential to 
study the unusual big losses in insurance, analyze equity risk, predict rare natural disasters.
For all those cases, probability of the event of interest is very small compared to our past observations
although very crucial as they may produce a huge risk (loss) in practice; 
known as the ``worst-case risk".  
The statistical tool that helps to analyze such situations is the extreme value theory.
Thus, in recent era, there is a growing trend of fruitful literatures about the inference 
under extreme value models. These models generally have a thicker tail compared to the normal models
and the probabilities of rare events are modeled by non-zero tail probabilities
of such heavy-tailed distributions.   For univariate cases, such distributions can be characterized 
by its tail index that measure, in layman's term, the thickness of the tail and by some assumptions about 
its probability density function in tail region. For multivariate cases, although the marginals can be
characterized in the same way, specification of their joint distribution requires additional effort 
to capture their dependence structure in the tail. Further, this tail dependence always may not have 
a linear structure; for example, in financial markets, the dependences between returns are mostly 
non-linear. Generally such non-linear dependence in the tail region is modeled by means 
of ``Copula" function specifying its pattern and the extend of dependence is measured 
by introducing an index parameter, known as the coefficient of tail dependence. 
Thus, the estimation of this index parameter is crucial in multivariate extreme value theory 
and, as one can expect, there exist several estimators for this purpose; for example see the works of 
Ledford and Tawn (1996, 1997), Beirlant and Vandewalle (2002), Heffernan and Tawn (2004), 
Draisma et al.~(2004), Goegebeur and Guillou (2012) among others.

However, existing literatures do not take care into account the outlying observations 
present in data and most of the estimators of tail dependence coefficient, if not all, 
are highly sensitive to those outliers. However, in real practice, there can be a significant portion 
of outliers in sample with respect to the assumed model, 
either due to ignorance of some external factor, or erroneous input in some level of data collection; 
but ignoring the external factors are not recommended 
while we are using a bivariate model to  asses risk or market returns.
In such cases, our inference without proper control over these outliers about the tail events may 
changes drastically and overall objective of the study fails to produce a big loss in most cases.
There are some recent attempts to produce robust estimator of the tail index in univariate 
cases (see Kim and Lee, 2008, for example); 
but the multivariate counterpart seem not to receive such attention till now. 
In this paper, we consider the bivariate extreme value distribution and propose some 
estimators for the coefficient of the tail dependence that are highly robust with respect to the 
outlying observations in the sample; also the proposed estimators will be seen to give 
quite competitive and sometimes better results even in case of pure data. 
However, to make the paper appealing to a large class of applied scientist even outside the community 
of  mathematicians, we will not present any complicated asymptotics of the proposed estimators; 
rather illustrate their robustness through the graphical approach of influence function analysis and 
their bias and MSE through various numerical examples considering several kind of dependence structures
and presence of possible outliers.

The rest of the paper is organized as follows: We start with a brief introduction to the 
bivariate heavy tailed distributions, its coefficient of tail dependence and 
some common existing estimators in Section \ref{SEC:HT_distr_intro}.
In section \ref{SEC:robust_est} we will propose some new estimator of the tail dependence coefficient
with special attention to its robustness and explore their robustness using 
the classical influence function analysis in Section \ref{SEC:robustness}.
The performance of the proposed estimators will be examined though an extensive simulation study in 
Section \ref{SEC:numerical}. Finally we will end this paper with a short overall discussion.

\section{Bivariate Heavy Tailed Distributions and Estimation of Tail Dependence Coefficient}
\label{SEC:HT_distr_intro}

Suppose $(X,~Y)$ denotes a pair of random variables with joint distribution function $F$
and marginal distribution functions $F_X$ and $F_Y$ respectively.
Further let the marginals to be known having unit Fr\'{e}chet distributions ($e^{-\frac{1}{u}}$ for $u>0$) 
or unit Pareto distribution (${1-\frac{1}{u}}$ for $u>1$). 
In most practical cases, estimation of probability of any tail event was done based on 
the assumption of asymptotic dependence between the two variables as 
it was difficult to obtain the limiting joint probability estimate of any extreme event 
(both variables exceeding some large threshold depending on the sample size). 
The problem was solved by Ledford and Tawn (1996, 1997) who proposed an alternative model 
for joint probability of bivariate extreme event that can be estimated even under asymptotic independence.
They have assumed that the probability $P(X>t, ~Y>t)$ is a regularly varying function of $t$ at zero 
with index $\frac{1}{\eta}$; more precisely, for $\eta \in (0,~1]$
\begin{equation}
P(X>t, ~Y>t) \sim \mathcal{L}(t) t^{-\frac{1}{\eta}} ~~~~~~\mbox{as }~t \rightarrow \infty,
\label{EQ:HT_def}
\end{equation}  
where $\mathcal{L}$ is a slowly varying function satisfying 
$
\frac{\mathcal{L}(tx)}{\mathcal{L}(t)} \rightarrow 1 ~~~~~~\mbox{as }~t \rightarrow \infty, ~~~ \forall~~x>0.
$
The parameter $\eta$ measures the extent of asymptotic tail dependence between the two variables $(X,~Y)$
with larger values indicating greater association 
and is known as the coefficient of tail dependence or the tail dependence index.
In particular, the cases with $\eta=1$ and $\lim\limits_{t\rightarrow~0}\mathcal{L}(t) = c \in (0,~1]$ 
represent asymptotic dependence whereas all other cases represent asymptotic independence. 
Ledford and Tawn (1996) further identified three types of asymptotic independence based on the 
values of $\eta$; see also Heffernan (2000) or Goegebeur and Guillou (2012).

Further note that, under the assumption of continuous marginals (either Fr\'{e}chet or Pareto), we have 
$\mathbb{P}\big(1-F_X(X)<tx, 1-F_Y(Y)<ty\big) = tx + ty -1 + C(1-tx, 1-ty),$
with $C$ being the copula function of joint distribution $F$ and hence all the bivariate distributions 
can be characterized in terms of their copula function. 
Some useful examples of copula structures are given below (see, e.g., Aas, 2004 for details):

\begin{itemize}
\item \textbf{Bivariate Normal Copula:~~}
It corresponds to the most popular bivariate Normal distribution (symmetric)
with correlation coefficient $\rho \in [-1,~1]$. The case with  $|\rho| < 1$ have asymptotically 
independent structure 
with $\eta=\frac{(1+\rho)}{2}$.
Thus, in general, this simple model does not allow for joint fat-tail.
\\
  
\item \textbf{Bivariate Student's $t$-Copula:~~}
It corresponds to the bivariate student's $t$-distribution (symmetric)
with correlation coefficient $\rho \in [-1,~1]$ and degrees of freedom $\nu$. 
This model allows for joint fat-tail and higher probability of co-extreme events regardless of 
their marginal distributions; however, the power of exhibiting such events decreases as $\nu$ increases. 
See Demarta and McNeil (2005) for more details on $t$-Copula and the formuls for its tail dependence coefficient.
\\

\item \textbf{Gumbel copula:~~}  
It is an asymmetric copula allowing us to model the same in dependence structures, 
for example the equity return dependence structures reported by Longin and Solnik (2001) or 
Ang and Chen (2000). It exhibits greater dependence in the positive tail and has an explicit form given by 
$
C_\delta(u, v) = exp\left( - [(-\log~u)^\delta + (-\log~v)^\delta]^{1/\delta}\right),
$
Here $\delta \in (0,~1]$ controls the extent of dependence; 
$\delta \rightarrow 0$ gives perfect dependence and $\delta=1$ corresponds to independence with $\eta=0.5$.
\\

\item \textbf{Clayton copula:~~}
It is also an asymmetric copula, but exhibits greater dependence in the negative tail.  
It has the explicit form given by 
$
C_\delta(u, v) = \left( u^{-\delta} + v^{-\delta} - 1 \right)^{1/\delta},
$
where $\delta \in (0,~\infty)$ controls the extent of dependence; 
$\delta \rightarrow 0$ implies independence and 
$\delta \rightarrow \infty$ corresponds to perfect dependence.
Here also, $\delta=1$ yields $\eta =0.5$.

\end{itemize}

Our main objective is to estimate the dependence index $\eta$ based on a observed sample $(X_i,~Y_i)$
of size $n$ on the random vector $(X,~Y)$ and there exist different methods for this estimation problem.
The main root to most of these methods is to transfer the problem into one-dimensional tail index estimation
and then apply the suitable techniques on the transformed univariate heavy-tailed random variable.
Let us assume $Z_i ={\rm min}\{X_i, ~Y_i\}$ for all $i=1, \ldots, n$.  Then
\begin{equation}
P(Z_i>t) = P(X_i>t, ~Y_i>t) \sim \mathcal{L}(t) t^{-\frac{1}{\eta}} ~~~~~~\mbox{as }~t \rightarrow \infty,
\label{EQ:transformed_univariate}
\end{equation}
so that $Z_1, \ldots, Z_n$ represent independent and identically distributed random observations
from  a univariate heavy-tailed distribution with tail index $\eta$. 
Therefore, one can estimate the dependence index $\eta$ by applying 
the univariate theory on the minimum of two variables; 
but in that case we must standardized the marginal distribution first to the assumed one
--- unit  Fr\'{e}chet or unit Pareto distribution. 
For the observed sample, this standardization can be done using their empirical marginals; 
let $R(X_i)$ and $R(Y_i)$ denote the rank of $X_i$ and $Y_i$ respectively in the sample.
Then we use minimum of the unit  Fr\'{e}chet transformed marginals given by 
\begin{equation}
Z_i = \min\left\{-\frac{1}{\log(R(X_i)/(n+1))}, ~ -\frac{1}{\log(R(Y_i)/(n+1))} \right\}, ~~~ i=1,\ldots,n,
\label{EQ:Frechet_transform}
\end{equation}
or use minimum of the unit  Pareto transformed marginals given by 
\begin{equation}
Z_i = \min\left\{\frac{1}{1 -(R(X_i)/(n+1))}, ~ \frac{1}{1-(R(Y_i)/(n+1))} \right\}, ~~~ i=1,\ldots,n.
\label{EQ:pareto_transform}
\end{equation}
Now, there are many techniques that can be applied to this univariate problem 
and their properties are discussed in the context of estimating the bivariate tail 
dependence index by many researchers; see Hill (1975), Smith (1987), Dekkers et al.~(1989), 
Beirlant and Vandewalle (2002), Beirlant et al.~(2011), Goegebeur and Guillou (2012) among others.
Among  all these estimators, the simplest and classical technique is the Hill estimator, 
which is basically the maximum likelihood estimators assuming the exponential distribution 
for the logarithmic relative excesses over a given threshold; it also has an explicit form given by
\begin{equation}
\widehat{\eta}_H = \frac{1}{k} ~\sum_{i=1}^k\log(Z_{(n-i+1)}) - \log(Z_{(n-k)}),
\label{EQ:Hill_est}
\end{equation}
where $Z_{(i)}$ denotes the $i^{\rm th}$ order statistics in $\{Z_1, \ldots, Z_n\}$.
The estimator used by  Beirlant and Vandewalle (2002) is also a maximum likelihood estimator based on 
an exponential regression model approximation. Alternatively Dekkers et al.~(1989)
used a moment estimator and Goegebeur and Guillou (2012) used a functional estimator of $\eta$.
However, these estimators along with most of the others are non-robust with respect to 
the presence of outliers in data. The non-robust nature of similar estimators was observed in case of 
univariate models by many researchers and there is few recent attempts to make them robust. 
Here we will consider two such robust methods that exploits the special structure of 
the density power divergence (Basu et al., 1998) and down-weights the outlying observations 
by a non-zero power of model density to generate robust inference.
One such approach is introduced by Kim and Lee (2008) 
who have generalized the Hill's estimator to achieve robustness under similar model assumptions; 
another proposal has been made very recently by Ghosh (2014) assuming the exponential regression model 
and hence produce robust generalization of the Beirlant and Vandewalle (2002) estimator. 
We will apply these two concepts in the present bivariate problem of estimating the coefficient of 
tail dependence and study their properties in view of several bivariate heavy-tailed model.

\section{Robust Estimators of Tail Dependence Coefficients}
\label{SEC:robust_est}

%
%
%

\subsection{Estimation under Exponential Log-Relative Excess}

Consider the set-up of $n$ bivariate heavy-tailed observations with unit Fr\'{e}chet marginals 
and define the transformed variables $Z_i$ by (\ref{EQ:Frechet_transform}). 
We will apply the robust generalization of the Hill's estimator to this case following Kim and Lee (2008). 
Let $F_Z$ denotes the distribution function of the unit Fr\'{e}chet transformed marginals $Z_i$
which, in view of (\ref{EQ:transformed_univariate}), is a heavy-tailed distribution with tail index $\eta$.
We will assume that the log-relative excess of the variable $\{Z_i\}$ over a given threshold follows 
an exponential distribution with mean $\eta$, i.e., for all $z>0$ there exists a sequence $\{k_n\}$ 
of positive numbers satisfying $k_n \rightarrow\infty$, $\frac{k_n}{n}\rightarrow 0$ and
\begin{equation}
\label{EQ:exp_assump}
\frac{1}{k_n} \sum_{i=1}^n ~ I\left(\log Z_i - \log b(n/k_n) > z\right) 
\mathop{\rightarrow}^\mathcal{P} e^{-\frac{z}{\eta}} ~~~~~\mbox{as }~n\rightarrow \infty, 
\end{equation}
where $b(z) = F_Z^{-1}(1-\frac{1}{z})$. Under this assumption, 
we can model $\widetilde{Z}_i = \log Z_i - \log b(n/k_n)$ 
as $n$ independent and identically distributed observation from exponential distribution with mean $\eta$. 
Then $\eta$ can be estimated by minimizing the density power divergence (DPD) between 
the data and the exponential model (Basu et al., 1998). The density power divergence down-weights the outlying
observations by $\alpha^{\rm th}$ power of model density with $\alpha \in [0,1]$ and generate robust estimators;
note that the case $\alpha=0$ produces no outlier down-weighting at all and 
coincides with the non-robust maximum likelihood estimator.
Further, we will replace the unknown term $b(n/k_n)$ by $Z_{(n-k)}$ following the suggestion of Kim and Lee (2008).
Then, a routine calculation shows that the corresponding estimating equation is given by 
\begin{equation}
\frac{\alpha}{(1+\alpha)^2\eta} + \frac{1}{k}\sum_{i=1}^n 
\left(\frac{\widetilde{Z}_i}{\eta^2} - \frac{1}{\eta}\right) e^{-\frac{\alpha\widetilde{Z}_i}{\eta}} 
I(\widetilde{Z}_i>0) = 0, \label{EQ:est_eqn_DPD}
\end{equation}
where $k=k_n$ and $\widetilde{Z}_i = \log Z_i - \log Z_{(n-k)}$. 
The solution of this estimating equation will give us an estimator of $\eta$ that we call as the DPD based 
estimator of $\eta$ and denote it by $\hat{\eta}_D^\alpha$. Clearly, for any fixed sample size $n$, 
this estimator depends on the choice of tuning parameters $\alpha$ and $k =k_n$.
Interestingly, the estimator $\hat{\eta}_D^0$ coincides with the Hill's estimator defined in (\ref{EQ:Hill_est}).

%

\subsection{Estimation under Exponential Regression Model}
\label{SEC:robust_est_ERP}

As an alternative to the above approach of robust estimation of $\eta$, 
we will now present another type of estimators with the exponential regression model approximation 
as given in Matthyas and Beirlant (2001). 
Again consider the set-up of $n$ bivariate heavy-tailed observations as given in 
Section \ref{SEC:HT_distr_intro} with the transformed univariate observations $\{Z_i\}$. 
Assume that the scaled log-ratios of excess in the univariate heavy-tailed observations $Z_i$  
over a large threshold $Z_{(n-k)}$, defined as 
$$
W_j = j \log\left(\frac{Z_{(n-j+1)} - Z_{(n-k)}}{Z_{(n-j)} - Z_{(n-k)}}\right), ~~~~ j = 1, 2, \ldots, k-1,
$$
follow approximately an exponential regression model (ERM) 
\begin{equation}
W_j \mathop{\approx}^\mathcal{D} \theta_j E_j,  ~~~~ j = 1, 2, \ldots, k-1,
\label{EQ:ERM_asump}
\end{equation}
with $\theta_j = \frac{\eta}{1 - (j/k)^\eta}$ and 
$E_j$ being independent and identically distributed random variables with mean $1$.

Beirlant and Vandewalle (2002) used this approximation and obtained the estimator of 
tail dependence coefficient $\eta$ using the maximum likelihood approach, 
which  has the well-known lack of robustness property.
Note that, the random variables $W_1, \ldots, W_{k-1}$ are 
independent but not identically distributed under above approximation. 
Recently Ghosh and Basu (2013) developed a robust estimation procedure for 
such cases of independent but non-homogeneous observations using the density power divergence 
and Ghosh (2014) applied the same for estimating the tail index 
in univariate case with the exponential regression model. 
Here, to estimate the tail dependence index $\eta$ under bivariate models, we can use the similar approach;
obtain the estimator  minimizing the average density power divergence over those $k-1$ non-homogeneous data points.
As noted in Beirlant and Vandewalle (2002), with this approximation of ERM, 
we can assume both  Fr\'{e}chet or Pareto distribution for marginals and define the transformed 
variables $Z_i$ accordingly using (\ref{EQ:Frechet_transform}) or (\ref{EQ:pareto_transform}) respectively.
Then, the estimating equation of this case is given by 
\begin{equation}
\sum_{j=1}^{k-1}  \widetilde{J}_\alpha\left(\frac{j}{k+1}\right)\left[\frac{\alpha\theta_j}{(1+\alpha)^2} + 
\left({W_j} - {\theta_j}\right) e^{-\frac{\alpha W_j}{\theta_j}} \right] = 0, 
\label{EQ:est_eqn_ERM}
\end{equation}
where $\widetilde{J}_\alpha(u) = (u^\eta - 1 - \eta u^\eta \log u)(1-u^\eta)^\alpha \eta ^{-\alpha-2}$.
We will denote the estimator obtained as a solution to this estimating equation  as $\hat{\eta}_{ER,F}^\alpha$
and $\hat{\eta}_{ER,P}^\alpha$ respectively based on whether the marginals are assumed to be Fr\'{e}chet or Pareto
distributed. Note that, the case with $\alpha=0$ coincides with the estimator obtained by 
Beirlant and Vandewalle (2002) and the estimator with $\alpha>0$ produces a robust generalization of 
their estimator of tail dependence coefficient.

%

\section{Robustness Analysis}
\label{SEC:robustness}

\subsection{Influence Function and its Interpretation}
\label{SEC:IF}

The influence function introduced by Hampel (1968, 1974) is a classical tool used in robust statistics; 
it helps to examine the behavior of an estimator under infinitesimal contamination 
through corresponding statistical functional.
Consider a random sample $U_1, \ldots, U_n$ from a $p$-variate distribution $G$ which 
we want to model by a parametric family $\{F_\theta: \theta \in \Theta \subseteq \mathbb{R}^d\}$. 
Suppose our estimator $\hat{\theta}$ of $\theta$ can be expressed as a solution of the estimating equation 
\begin{eqnarray}
\sum_{i=1}^n ~\psi(U_i, \theta) = 0,
\label{EQ:M-est}
\end{eqnarray}
for suitable function $\psi(\cdot,\cdot)$; such estimators with a few additional conditions are 
known as the M-estimator and can be robust with some particular choices of $\psi$.
Let $G_n$ denotes the empirical distribution function based on the above sample.
Then, we can write the estimator as $\hat{\theta}=T(G_n)$ for the corresponding statistical functional $T(G)$, 
defined as the solution of 
\begin{eqnarray}
\int ~\psi(x, \theta) dG(x) = 0.
\label{EQ:IF-est_eqn}
\end{eqnarray}
Note that we get equation (\ref{EQ:M-est}) from (\ref{EQ:IF-est_eqn}) replacing $G$ by $G_n$.
The value of $\theta^g = T(G)$ is called the best fitting parameter under the true distribution $G$;
if the true distribution belongs to the model family with $G=F_{\theta_0}$, 
then we should have $\theta^g =\theta_0$.
This philosophy is used for all the minimum distance (or divergence) estimators 
including the maximum likelihood estimator.

Consider the contaminated distribution $G_\epsilon = (1-\epsilon) G + \epsilon \wedge_y$, 
where $\wedge_y$ is the degenerate distribution at the contamination point $y$ 
(in the sample space) and let $\theta_\epsilon = T(G_\epsilon)$. 
Then the influence function of the estimator defined by the functional $T(\cdot)$ 
at the true distribution $G$ is defined as 
$$
IF(y; T, G) = \lim\limits_{\epsilon \downarrow 0} ~\left[\frac{T(G_\epsilon) - T(G)}{\epsilon}\right]
= \left.\frac{\partial \theta_\epsilon}{\partial \epsilon}\right|_{\epsilon=0},
$$
provided it exists. Thus the influence function measures the effect of infinitesimal contamination 
at a point $y$ on the asymptotic bias of the estimator, standardized by the by the proportion of contamination.

To obtain the influence function for the functional estimator $T_\psi$ defined by (\ref{EQ:IF-est_eqn}),
we replace $\theta$ and $G$ in (\ref{EQ:IF-est_eqn}) by $\theta_\epsilon$ and $G_\epsilon$ respectively
and then differentiate it with respect to $\epsilon$ at $\epsilon=0$; 
after collecting suitable terms it yields the required influence function having the form
\begin{equation}
IF(y; T_\psi, G) = \left(\int ~\nabla\psi(x, \theta) dG(x)\right)^{-1}\psi(y, \theta^g).
\label{EQ:IF_M-est}
\end{equation}
Note that this influence function is bounded with respect to $y$ 
whenever the term $\psi(y, \theta)$ is bounded with respect to $y$ for all $\theta$; 
this plays the main role to choose suitable $\psi$-function generating robust M-estimators.

\subsection{Influence Function of $\hat{\eta}_D^\alpha$}
\label{SEC:IF_DPD}

Now let us derive the influence function of the proposed DPD based estimator $\hat{\eta}_D^\alpha$
to get the idea about its robustness. Note that the corresponding estimating equation (\ref{EQ:est_eqn_DPD}) is 
of the form (\ref{EQ:M-est}) with $\widetilde{Z}_i = \log Z_i - \log b(n/k_n)$ being i.i.d.~from 
their true distribution function $G$, $\theta=\eta$ and
\begin{equation}
\psi(x, \theta) = \left(\frac{x}{\theta^2} - \frac{1}{\theta}\right)e^{-\alpha\frac{x}{\theta}} 
+ \frac{\alpha}{(1+\alpha)^2\theta} ;
\label{EQ:psi_func}
\end{equation}
so we can use the technique described above to obtain its influence function. 
Note that if $F(x,y)$ is the true distribution of the bivariate heavy-tailed random variables $(X,~Y)$ 
with marginals $F_x$ and $F_y$, then the distribution of $\widetilde{Z}_i $  is given by 
$G(z) = F_X(be^z) + F_Y(be^z) - F(b(n/k_n)e^z,b(n/k_n)e^z)$. 
Define corresponding statistical functional $T_D^\alpha(G)$ at this $G$,
as solution of the equation 
$\int ~\psi(z, \theta) dG(z) = 0.$
Consider contamination on the bivariate distribution $F(x,y)$ at some point $(t_1, t_2)$ and define
$F_\epsilon(x,y) = (1-\epsilon)F(x,y) + \epsilon \wedge_{(t_1, t_2)}$. 
Then it induces contamination on the distribution of $\widetilde{Z}_i$ as 
$G_\epsilon(z) = (1-\epsilon)G(z) + \epsilon \wedge_{b(n/k_n)\min(e^{t_1}, e^{t_2})}$. 
Using this contaminated distribution $G_\epsilon$ and proceeding as in subsection \ref{SEC:IF}, 
or using equation (\ref{EQ:IF_M-est}), we get the influence function of the proposed estimator 
$\hat{\eta}_D^\alpha$ at the true distribution $F$ given by  
\begin{eqnarray}
&& IF((t_1, t_2); T_D^\alpha, F) \nonumber\\ 
&& = \left[\int~\left(\frac{(1+\alpha) z}{\eta} - \frac{\alpha z^2}{\eta^2}\right) 
e^{-\frac{\alpha z}{\eta}} dG(z) \right]^{-1}  \nonumber\\
&& \times \left[\left(b(n/k_n)\min(e^{t_1}, e^{t_2})-\eta\right) 
e^{-\alpha\frac{b(n/k_n)\min(e^{t_1}, e^{t_2})}{\eta}}
 -\int~(z-\eta)e^{-\frac{\alpha z}{\eta}}dG(z)\right]. \nonumber\\
\label{EQ:IF_DPD-est}
\end{eqnarray}

Now consider the particular case of true distributions for which 
the assumption (\ref{EQ:exp_assump}) holds true so that we can approximate $G(z)$ by $ 1 - e^{-\frac{z}{\eta}}$; 
in that case the influence function of  $\hat{\eta}_D^\alpha$  can be simplified to have the form 
\begin{eqnarray}
&& IF((t_1, t_2); T_D^\alpha, F) \nonumber\\
&& = \frac{(1+\alpha)^3}{(1+\alpha^2)}
\left[\left(b(n/k_n)\min(e^{t_1}, e^{t_2})-\eta\right) e^{-\alpha\frac{b(n/k_n)\min(e^{t_1}, e^{t_2})}{\eta}} 
+ \frac{\alpha \eta}{(1+\alpha)^2}\right]. \nonumber\\
\label{EQ:IF0_DPD-est}
\end{eqnarray}
Clearly, this influence function is bounded for all $\alpha>0$ implying the robustness of 
corresponding estimators. However, at $\alpha=0$, the influence function of 
the corresponding estimator (Hill estimator) becomes 
$$
IF((t_1, t_2); T_D^0, F) = \left(b(n/k_n)\min(e^{t_1}, e^{t_2})-\eta\right),
$$
which is clearly unbounded with respect to the contamination point $(t_1, t_2)$. 
This proves the well-known non-robust nature of the Hill's estimator.
Figure \ref{FIG:IF_DPD5} below shows the plots of this influence function for different $\alpha$ 
at the true distribution satisfying (\ref{EQ:exp_assump}) with $\eta=0.5$ and 
$b(n/k_n) \approx 1 - \frac{k_n}{n} (k_n = n/2, n/4, n/10)$.
Note that, all influence function with $\alpha>0$ are bounded and their supremum decreases with increasing $\alpha$
which implies that the robustness of corresponding estimator increases as $\alpha$ increases.
However, their influence function and hence robustness does not differ significantly for the values of $k_n$.


\begin{figure}[!th]
\centering
\subfloat[$k_n = n/2$]{
\includegraphics[width=0.3\textwidth]{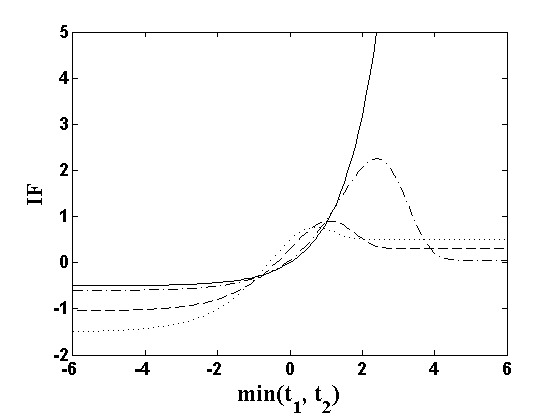}
\label{FIG:IF_DPD5_00}}
~ 
\subfloat[$k_n = n/4$]{
\includegraphics[width=0.3\textwidth]{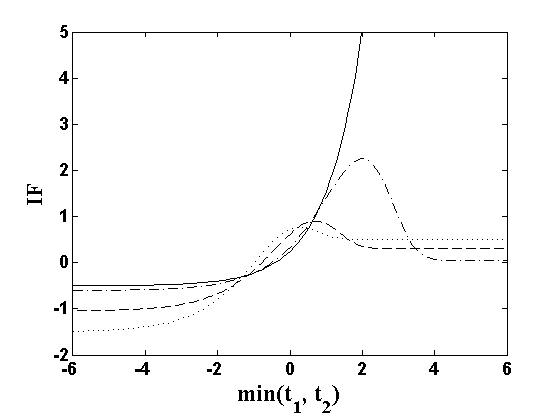}
\label{FIG:IF_DPD5_01}}
~
\subfloat[$k_n = n/10$]{
\includegraphics[width=0.3\textwidth]{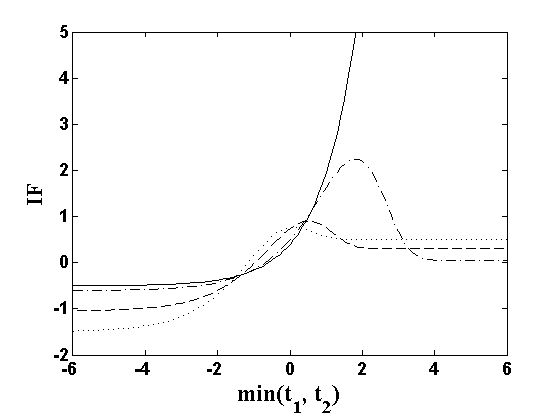}
\label{FIG:IF_DPD5_1}}
\caption{Influence function of $\hat{\eta}_D^\alpha$  for different $\alpha$ and $k_n$ at the true distribution with $\eta=0.5$ (solid line: $\alpha=0$, dashed-dotted line: $\alpha=0.1$, dashed line: $\alpha=0.5$, dotted line: $\alpha=1$).}
\label{FIG:IF_DPD5}
\end{figure}

\subsection{Influence Function of $\hat{\eta}_{ER,F}^\alpha$ and $\hat{\eta}_{ER,P}^\alpha$ }
\label{SEC:IF_ERM}

Now let us consider the alternative robust estimator $\hat{\eta}_{ER,F}^\alpha$ and $\hat{\eta}_{ER,P}^\alpha$ 
obtained under the assumption of the exponential regression model.
Note that, the corresponding estimating equations can also be written in form of equation (\ref{EQ:M-est})
with $\psi$ function defined by (\ref{EQ:psi_func}) and suitably defined $W_i$s, 
but here the random variables $W_i$s are not identically distributed. 
So, we can not obtain its influence function as defined in Section \ref{SEC:IF}.
In this case, we will follow Ghosh and Basu (2013) who have derived the influence function 
under the similar set-up of non-identically distributed observations; in contrast to the i.i.d.~case, 
their influence function and statistical functional under the non-homogeneous set-up depends 
on the sample size and is termed as the fixed-sample influence function. 
See Ghosh and Basu (2013) for more details.
Similar idea is used also by Ghosh (2014) in case of univariate tail index estimation under 
the similar exponential regression set-up.

Suppose the true distribution of the bivariate extreme value distribution is $F$ and 
in that case let the distribution of $W_j$ is denoted by $G_j$ for all $i=1, \ldots, k-1$; 
although all the $W_i$s are independent. Note that $G_j$s may depend on the marginal transformation used. 
In the estimation procedure we are approximating $G_j$ by an exponential distribution with mean $\theta_i$, 
as defined in Section \ref{SEC:robust_est_ERP}, for all $j$.
A contamination in the bivariate sample also produce a similar contamination on the transformed variables $W_j$s.
So, for simplicity, in this case we will work with $W_j$ and $G_j$s; so we will no need to consider 
the two cases of marginal transformation separately (as it will come through the construction of $G_j$s).
Let us define the corresponding statistical function $T_{ER,k}^\alpha$ for the exponential regression based 
estimator of tail dependence coefficient as the minimizer of 
$$
\sum_{j=1}^{k-1}~\int ~\psi(x, \theta) dG_j(x) = 0.
$$
Note that the above statistical functional depends on the  choice of the tuning parameter $k$ 
and so will be the influence function.

Now consider the contaminated distributions. 
Based on the amount of contamination in the original bivariate sample and the choice of $k$, 
there could be contamination in one or more of $W_j$s and on their distributions. 
First, let us assume contamination in only one $W_j$, say on $W_{j_0}$ for some $j_0 \in \{1, \ldots, k-1\}$
and let the contamination point be $t_0$.
Then, following the argument of Ghosh and Basu (2013), the corresponding influence function of the estimator 
$T_{ER,k}^\alpha$ at the true distribution $\underline{\mathbf{G}} = (G_1, \ldots, G_{k-1})$ is given by 
\begin{eqnarray}
&& IF(t_0; T_{ER,k}^\alpha, \underline{\mathbf{G}} ) \nonumber\\
&& = \frac{\Psi_n^{-1}}{k-1} 
\widetilde{J}_\alpha\left(\frac{j_0}{k+1}\right) \left[(t_0-\theta_{j_0})e^{-\alpha\frac{t_0}{\theta_{j_0}}}
-\int~(z-\theta_{j_0})e^{-\frac{\alpha z}{\theta_{j_0}}}dG_{j_0}(z)\right],~~~
\label{EQ:IF_ERM-est}
\end{eqnarray}
where $\Psi_n$ is defined according to Equations (3.3) and (3.5) of Ghosh and Basu (2013).
In particular, if the true distribution is such that the exponential regression model assumption 
(\ref{EQ:ERM_asump}) holds true, then the influence function of the estimator $T_{ER,k}^\alpha$ simplifies to 
\begin{eqnarray}
IF(t_0; T_{ER,k}^\alpha, F) &=& \frac{(1+\alpha)^3}{(1+\alpha^2)}  \left[\frac{1}{k-1}\sum_{j=1}^{k-1}
\frac{1}{\theta_j^{\alpha+2}}\widetilde{J}_0\left(\frac{j}{k+1}\right)\right]^{-1} \nonumber\\
&& ~~ \times\widetilde{J}_\alpha\left(\frac{j_0}{k+1}\right) \left[\left({t_0}-{\theta_{j_0}}\right) 
e^{-\alpha\frac{t_0}{\theta_{j_0}}}+ \frac{\alpha\theta_{j_0}}{(1+\alpha)^2}\right].~~
\label{EQ:IF0_ERM-est}
\end{eqnarray}
Note the similarity of this influence function with that of the DPD based estimator $T_D^\alpha$.
As before, the above influence function is bounded for all $\alpha>0$ and unbounded at $\alpha=0$.
This proves the robustness of the proposed estimators with $\alpha>0$ over the existing estimator of 
Matthyas and Beirlant (2001), corresponding to $\alpha=0$, 
under the exponential regression model approximation.  
Figure \ref{FIG:IF_ERM5} below shows the influence function of $T_{ER,k}^\alpha$ over $t_0$ for different $\alpha$ 
at the true distribution satisfying (\ref{EQ:ERM_asump}) with $\eta=0.5$ and different values of $k$ and $j_0$.
The boundedness of the influence function for all $\alpha>0$ is quite clear from the figure and 
its supremum decreases as $\alpha$ increases implying greater robustness. 
Further, this supremum of the influence function also decreases with increasing $k$ and $j_0$;
then the maximm possible extent of bias due to the contamination in data decreases and hence the bias-robustness 
of the proposed estimators increases as $k$ or $j_0$ increases.
This is quite intuitive from the view of robust statistical analysis; 
as $j_0$ or $k$ increases the effective sample size has to increase automatically and 
therefore the efect of any fixed amount of contamination becomes lesser on any inference drawn (statistically)
based on those data.

In case of contamination in more than one $W_j$s, the influence function can be derived in the similar way.
For example, if there is contamination in all the $W_j$s at the points $\mathbf{t} = (t_1, \ldots, t_{k-1})$ 
respectively, then the corresponding influence function of $T_{ER,k}^\alpha$ at the true distribution satisfying 
(\ref{EQ:ERM_asump}) is given by
\begin{eqnarray}
IF(\mathbf{t}; T_{ER,k}^\alpha, F) &=& \frac{(1+\alpha)^3}{(1+\alpha^2)} 
\left[\sum_{j=1}^{k-1}\frac{1}{\theta_j^{\alpha+2}}\widetilde{J}_0\left(\frac{j}{k+1}\right)\right]^{-1}\nonumber\\
&& ~~ \times\sum_{j=1}^{k-1}~\widetilde{J}_\alpha\left(\frac{j}{k+1}\right)
\left[\left({t_j}- {\theta_{j}}\right) e^{-\alpha\frac{t_j}{\theta_{j}}} 
+ \frac{\alpha\theta_{j}}{(1+\alpha)^2}\right].~~
\label{EQ:IF0_ERM-est_all}
\end{eqnarray}
The nature of boundedness of this influence function is again similar to the above case; 
it is unbounded at $\alpha=0$ and bounded at $\alpha >0$.

\begin{figure}[!th]
\centering
\subfloat[$k=50$, $j_0=1$]{
\includegraphics[width=0.3\textwidth]{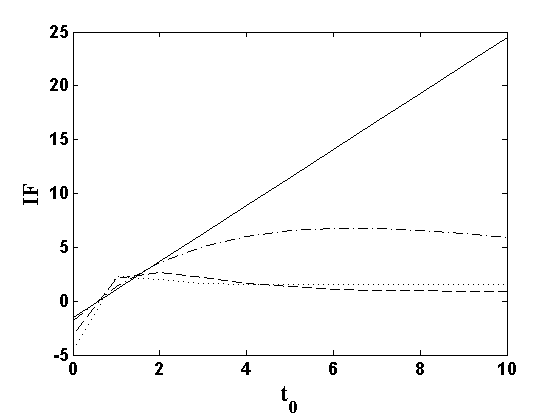}
\label{FIG:IF_ERM_k50_J1}}
~ 
\subfloat[$k=50$, $j_0=10$]{
\includegraphics[width=0.3\textwidth]{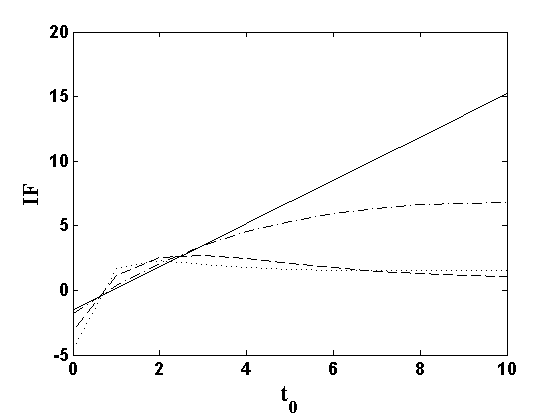}
\label{FIG:IF_ERM_k50_J10}}
~ 
\subfloat[$k=50$, $j_0=30$]{
\includegraphics[width=0.3\textwidth]{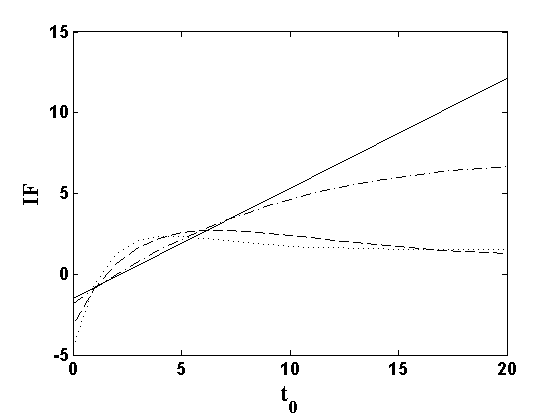}
\label{FIG:IF_ERM_k50_J30}}
\\
\subfloat[$k=100$, $j_0=1$]{
\includegraphics[width=0.3\textwidth]{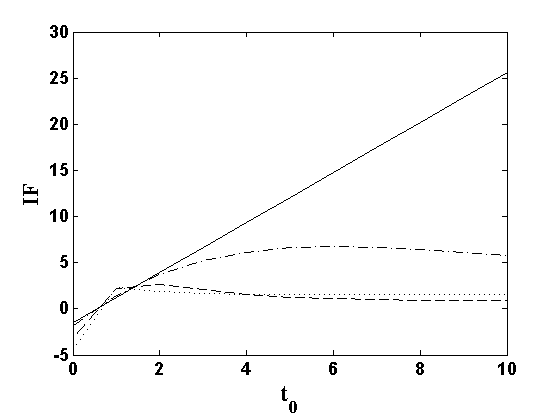}
\label{FIG:IF_ERM_k100_J1}}
~ 
\subfloat[$k=100$, $j_0=10$]{
\includegraphics[width=0.3\textwidth]{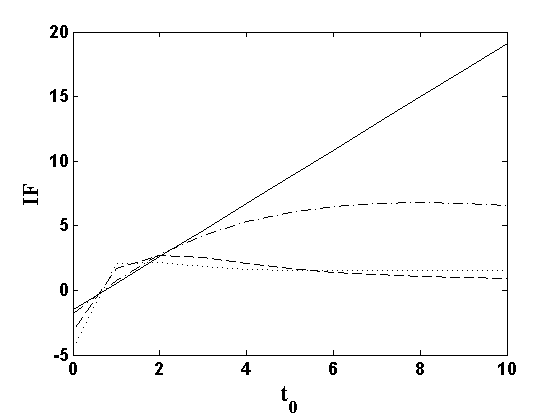}
\label{FIG:IF_ERM_k100_J10}}
~
\subfloat[$k=100$, $j_0=30$]{
\includegraphics[width=0.3\textwidth]{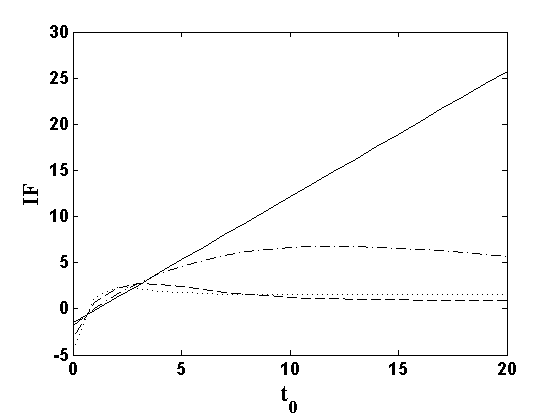}
\label{FIG:IF_ERM_k100_J30}}
\caption{Influence function of $T_{ER,k}^\alpha$  for different $\alpha$, $k$ and $j_0$ at the true distribution with $\eta=0.5$ (solid line: $\alpha=0$, dashed-dotted line: $\alpha=0.1$, dashed line: $\alpha=0.5$, dotted line: $\alpha=0$).}
\label{FIG:IF_ERM5}
\end{figure}

\section{Numerical Illustrations}
\label{SEC:numerical}

Let us now present the numerical illustration for the proposed robust estimators under several bivariate models.
First we examine their performance under the pure data with no contamination and 
see if we are loosing anything in order to obtain robustness under the contamination. 
We will consider the following important bivariate models for this purpose:

\begin{enumerate}
\item[(M1)] Bivariate normal distribution with both means $0$, both variances $1$ and correlation coefficient $0$.
This model has tail dependence coefficient $\eta=0.5$ and so it is asymptotically independent. 
Note that it is also stochastically independent.

\item[(M2)] Bivariate normal distribution with both means $0$, both variances $1$ and correlation coefficient $0.75$.
It has $\eta=0.875$ and is asymptotically independent 
with probability of co-extreme events being greater compared to that in stochastic independent case.

\item[(M3)] Standardized bivariate distribution generated by Gumbel Copula with parameter $\delta=1$ 
so that its tail dependence coefficient $\eta=0.5$.

\item[(M4)] Standardized bivariate distribution generated by Clayton Copula with parameter $\delta=1$ 
so that its tail dependence coefficient is again $\eta=0.5$.
\end{enumerate}
 
\begin{figure}[!th]
\centering
\includegraphics[width=0.24\textwidth]{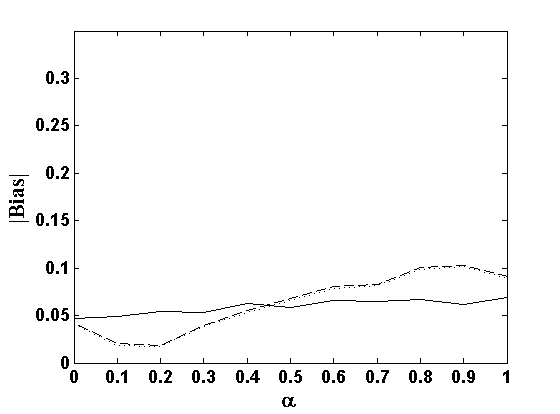}
\includegraphics[width=0.24\textwidth]{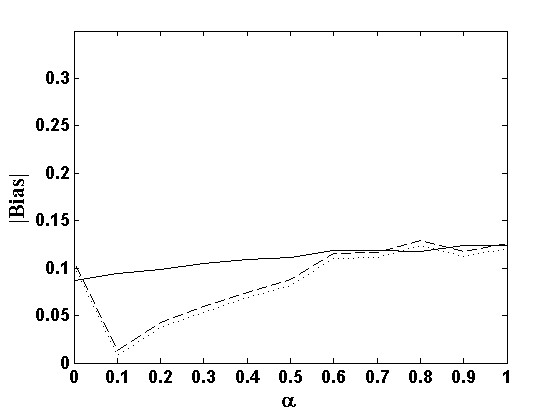}
\includegraphics[width=0.24\textwidth]{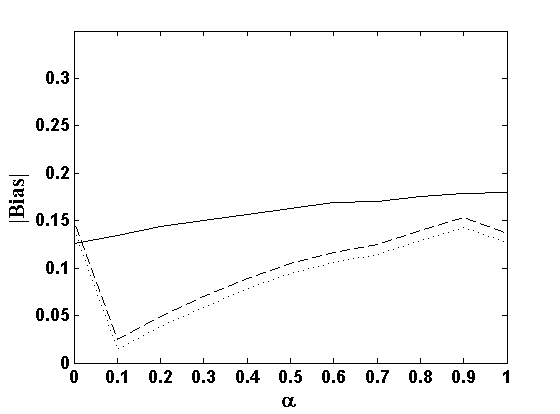}
\includegraphics[width=0.24\textwidth]{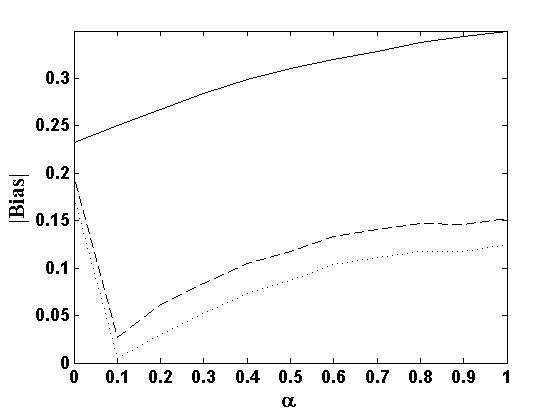}
\\
\subfloat[$k=50$]{
\includegraphics[width=0.24\textwidth]{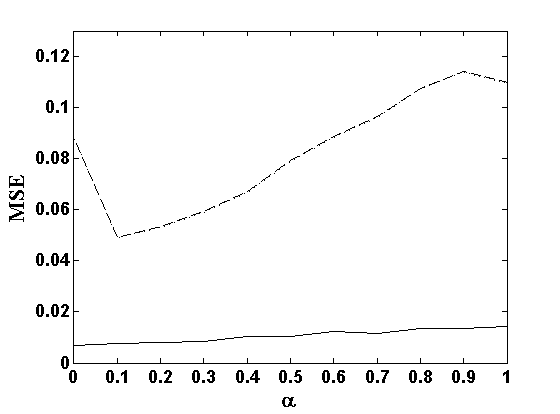}
\label{FIG:M1_k50}}
\subfloat[$k=150$]{
\includegraphics[width=0.24\textwidth]{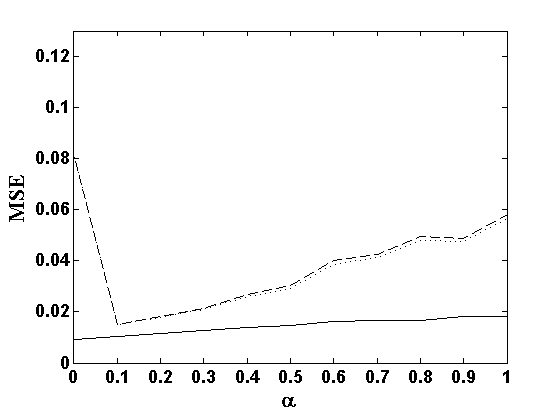}
\label{FIG:M1_k150}}
\subfloat[$k=250$]{
\includegraphics[width=0.24\textwidth]{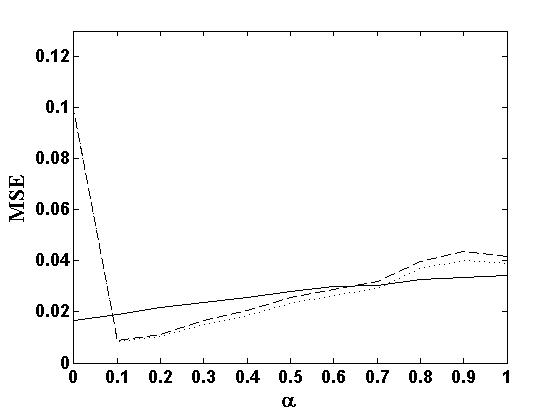}
\label{FIG:M1_k250}}
\subfloat[$k=500$]{
\includegraphics[width=0.24\textwidth]{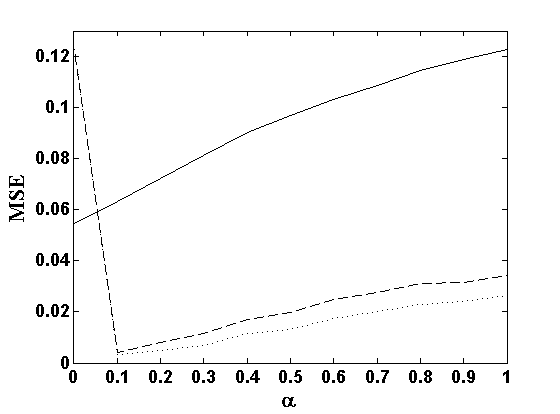}
\label{FIG:M1_k500}}
\caption{Empirical Bias and MSE of the estimators of $\eta$ for Model (M1) with $n=1000$. 
[solid line: $\hat{\eta}_D^\alpha$, dotted line: $\hat{\eta}_{ER,F}^\alpha$, dashed line: $\hat{\eta}_{ER,P}^\alpha$]}
\label{FIG:M1}
\end{figure}

\begin{figure}[!th]
\centering
\includegraphics[width=0.24\textwidth]{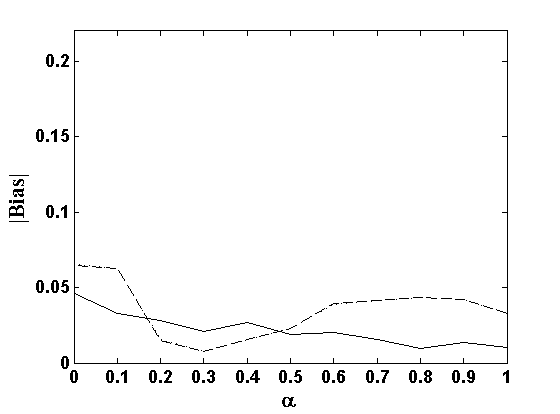}
\includegraphics[width=0.24\textwidth]{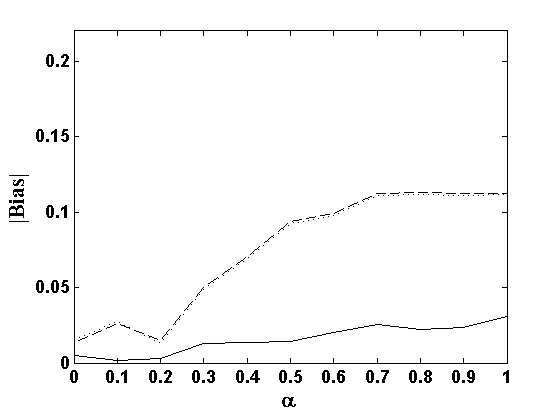}
\includegraphics[width=0.24\textwidth]{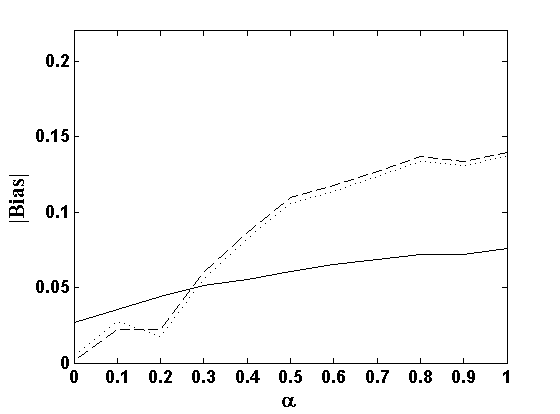}
\includegraphics[width=0.24\textwidth]{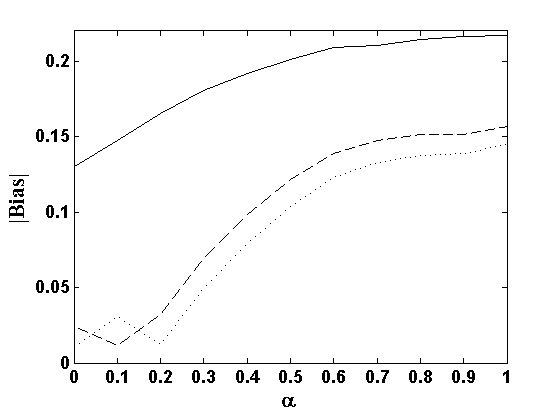}
\\
\subfloat[$k=50$]{
\includegraphics[width=0.24\textwidth]{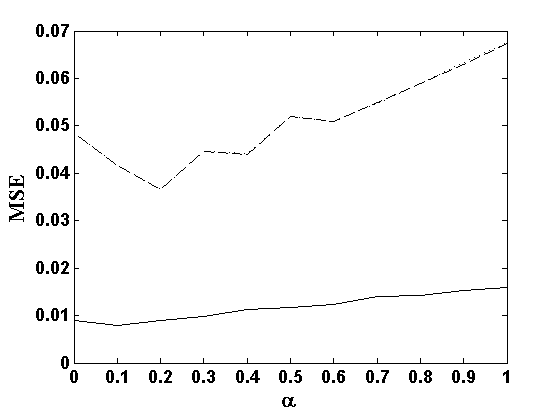}
\label{FIG:M2_k50}}
\subfloat[$k=150$]{
\includegraphics[width=0.24\textwidth]{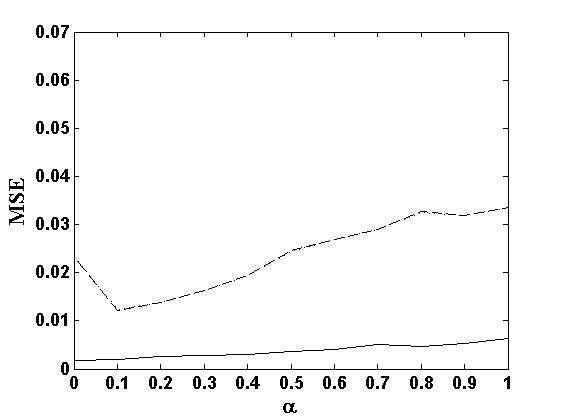}
\label{FIG:M2_k150}}
\subfloat[$k=250$]{
\includegraphics[width=0.24\textwidth]{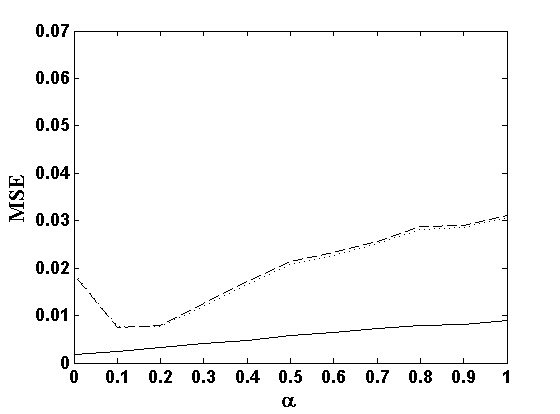}
\label{FIG:M2_k250}}
\subfloat[$k=500$]{
\includegraphics[width=0.24\textwidth]{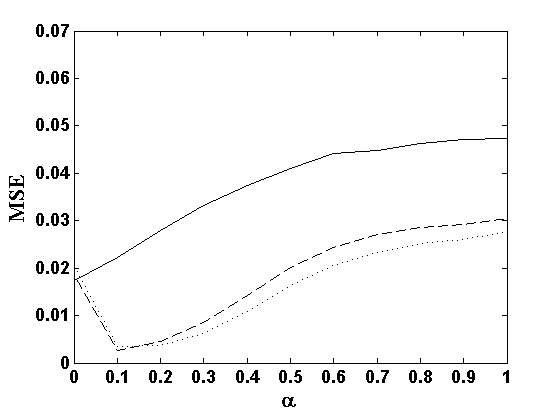}
\label{FIG:M2_k500}}
\caption{Empirical Bias and MSE of the estimators of $\eta$ for Model (M2) with $n=1000$. 
[solid line: $\hat{\eta}_D^\alpha$, dotted line: $\hat{\eta}_{ER,F}^\alpha$, dashed line: $\hat{\eta}_{ER,P}^\alpha$]}
\label{FIG:M2}
\end{figure}

\begin{figure}[!th]
\centering
\includegraphics[width=0.24\textwidth]{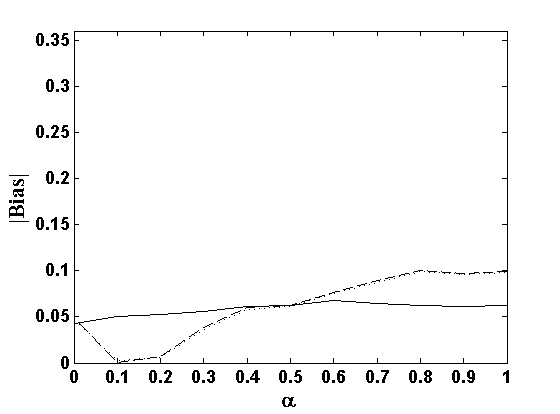}
\includegraphics[width=0.24\textwidth]{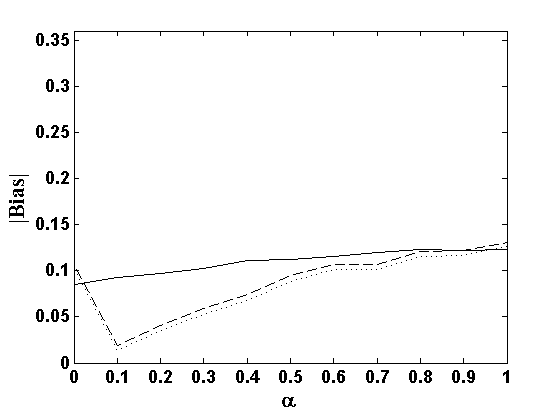}
\includegraphics[width=0.24\textwidth]{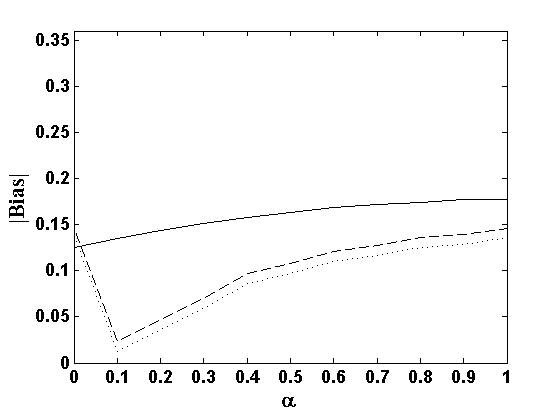}
\includegraphics[width=0.24\textwidth]{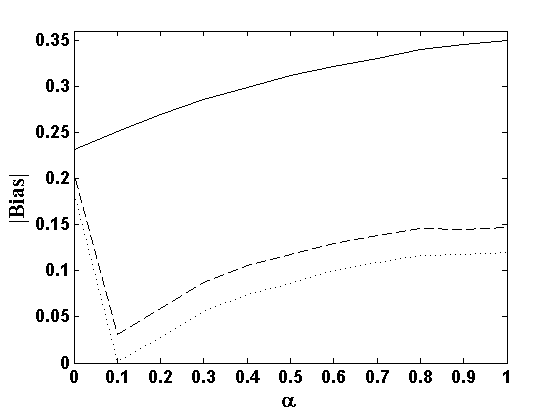}
\\
\subfloat[$k=50$]{
\includegraphics[width=0.24\textwidth]{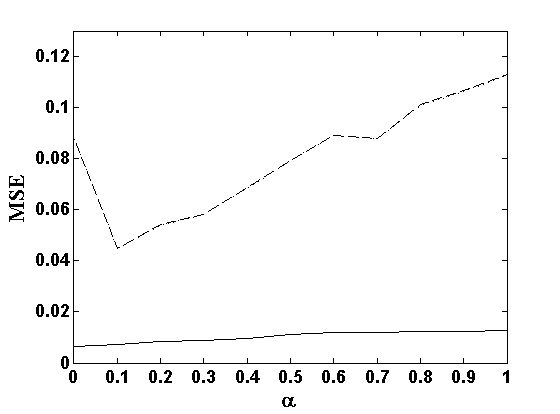}
\label{FIG:M3_k50}}
\subfloat[$k=150$]{
\includegraphics[width=0.24\textwidth]{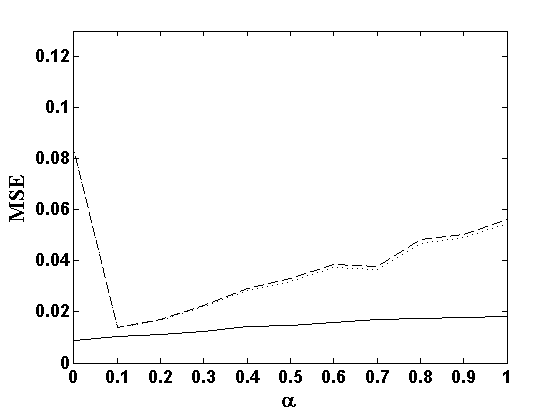}
\label{FIG:M3_k150}}
\subfloat[$k=250$]{
\includegraphics[width=0.24\textwidth]{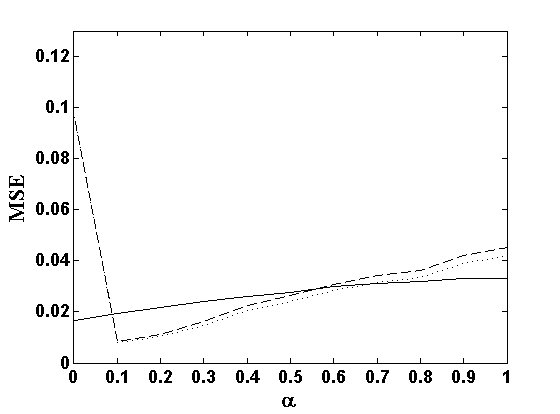}
\label{FIG:M3_k250}}
\subfloat[$k=500$]{
\includegraphics[width=0.24\textwidth]{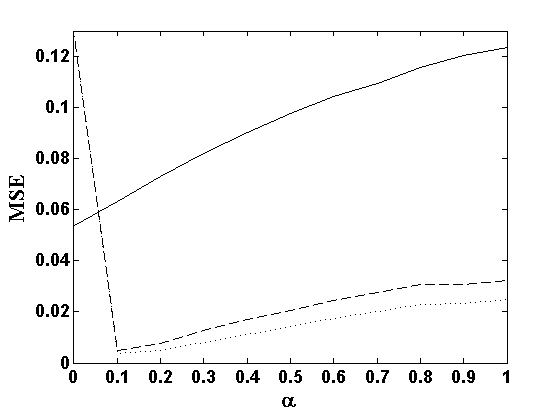}
\label{FIG:M3_k500}}
\caption{Empirical Bias and MSE of the estimators of $\eta$ for Model (M3) with $n=1000$. 
[solid line: $\hat{\eta}_D^\alpha$, dotted line: $\hat{\eta}_{ER,F}^\alpha$, dashed line: $\hat{\eta}_{ER,P}^\alpha$]}
\label{FIG:M3}
\end{figure}

\begin{figure}[!th]
\centering
\includegraphics[width=0.24\textwidth]{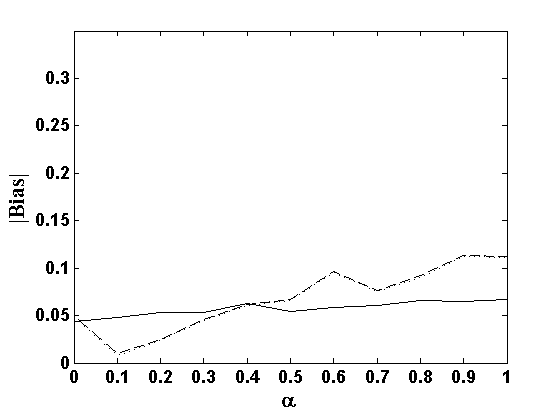}
\includegraphics[width=0.24\textwidth]{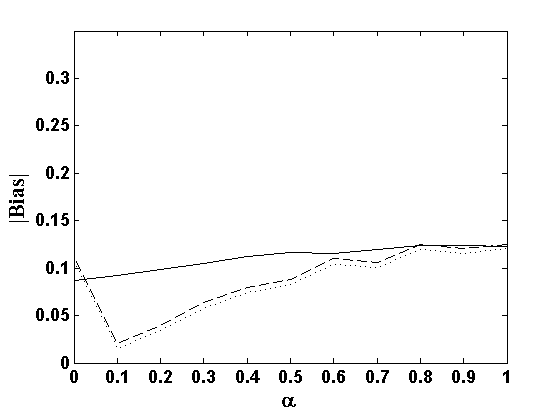}
\includegraphics[width=0.24\textwidth]{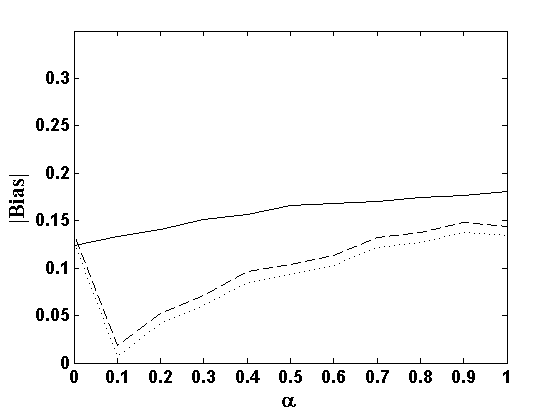}
\includegraphics[width=0.24\textwidth]{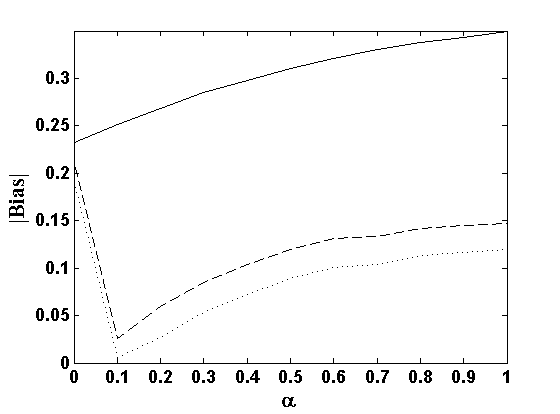}
\\
\subfloat[$k=50$]{
\includegraphics[width=0.24\textwidth]{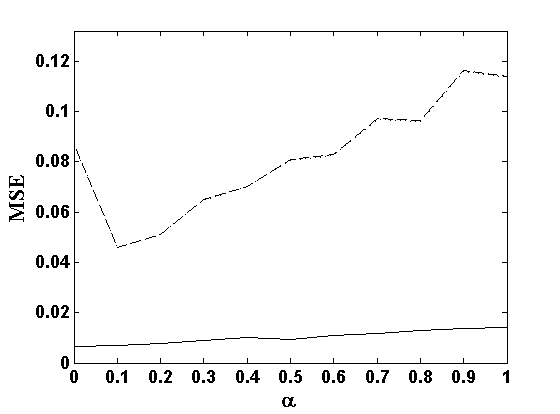}
\label{FIG:M4_k50}}
\subfloat[$k=150$]{
\includegraphics[width=0.24\textwidth]{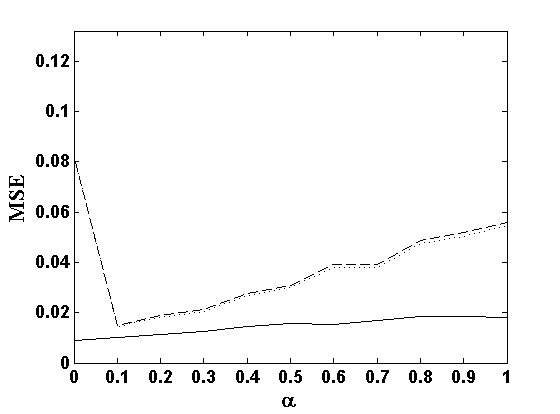}
\label{FIG:M4_k150}}
\subfloat[$k=250$]{
\includegraphics[width=0.24\textwidth]{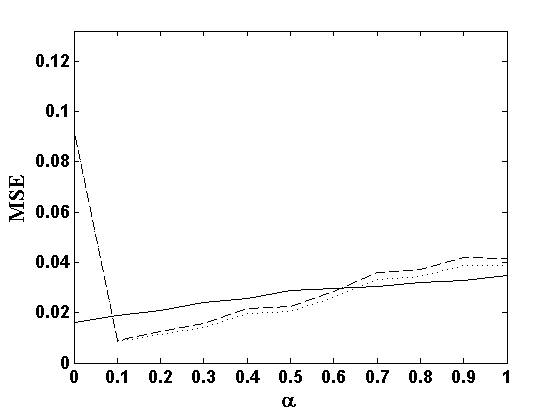}
\label{FIG:M4_k250}}
\subfloat[$k=500$]{
\includegraphics[width=0.24\textwidth]{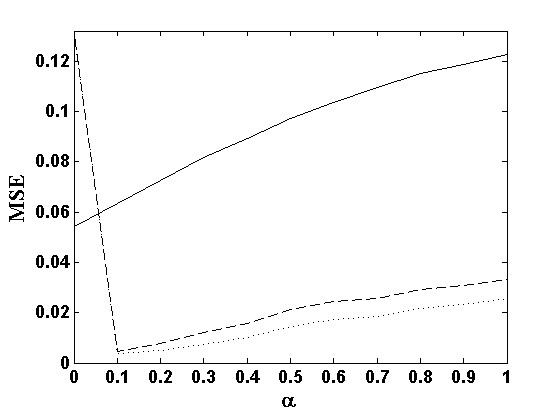}
\label{FIG:M4_k500}}
\caption{Empirical Bias and MSE of the estimators of $\eta$ for Model (M4) with $n=1000$. 
[solid line: $\hat{\eta}_D^\alpha$, dotted line: $\hat{\eta}_{ER,F}^\alpha$, dashed line: $\hat{\eta}_{ER,P}^\alpha$]}
\label{FIG:M4}
\end{figure}

For all the models, we have generated samples of size $n=500, 1000, 3000$ and 
estimated the coefficient of tail dependence using the proposed techniques, 
namely  $\hat{\eta}_D^\alpha$,$\hat{\eta}_{ER,F}^\alpha$ and $\hat{\eta}_{ER,P}^\alpha$ 
for several values of $\alpha$ and $k$. Then, we compute their empirical bias and MSE based on $1000$ replications
and compare their performances under all the models. 
However, for brevity, we only present the results for $n=1000$ in Figures \ref{FIG:M1} to \ref{FIG:M4};
the results of the cases $n=500$ and $n=3000$ are quite similar to this case and hence not reported here.
Further, it is noted that the performance of $\hat{\eta}_{ER,F}^\alpha$ and $\hat{\eta}_{ER,P}^\alpha$ 
are also quite similar in all the cases.

Analyzing the results for the model (M1), it can be seen that 
both bias and MSE of  $\hat{\eta}_D^\alpha$ increases slightly with $\alpha$ 
although the loss is not very significant;
their performance is also seen to be better for smaller values of $k$.
On the other hand, the bias and MSE of $\hat{\eta}_{ER,F}^\alpha$ first decreases with $\alpha$ and then increases;
the optimum values of $\alpha$ for which both are minimum lies near $0.1$. 
But, with respect to increasing values of $k$, its bias increases and MSE deceases; 
so we need to choose an optimum values of $k$.
It is seen that for model (M1) the optimum choice of parameters are $k \geq 250$ and $\alpha$ near $0.1$, 
which gives best result among those considered here and outperform the existing estimators -- 
Hill's estimator ($\hat{\eta}_{D}^\alpha$ with $\alpha=0$) and Matthyas and Beirlant (2001) 
estimator ($\hat{\eta}_{ER,F}^\alpha$ with $\alpha=0$) both in terms of bias and MSE; 
the extent of improvement increases as $k$ increases.
For the model (M2), the MSE of $\hat{\eta}_{D}^\alpha$ increases slightly with $\alpha$ 
but its bias decreases at small values of $k$. 
The performance of $\hat{\eta}_{ER,F}^\alpha$ is similar to (M1) but 
the optimum parameter is no longer at $k=250$ but more than that. 
At $k=500$ and $\alpha=0.1$, it shows huge improvement with respect to 
both bias and MSE over the existing estimators.   
Interestingly, the performance of the proposed estimator under models (M3) and (M4) are 
exactly the same as that under model (M1). 
This shows that the proposed estimators under pure data are also robust 
with respect to the model distribution but depends on the true parameter value $\eta$,
at least for the model considered here.

Next, to illustrate the robustness of the proposed estimators under contamination, 
we repeat the above simulation study but contaminate a certain proportion, say $\epsilon \%$ of the sample 
by outlying observations from some different distributions as follows:


\begin{figure}[!th]
\centering
\includegraphics[width=0.24\textwidth]{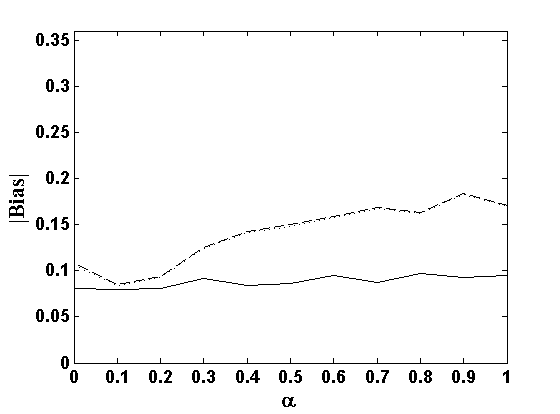}
\includegraphics[width=0.24\textwidth]{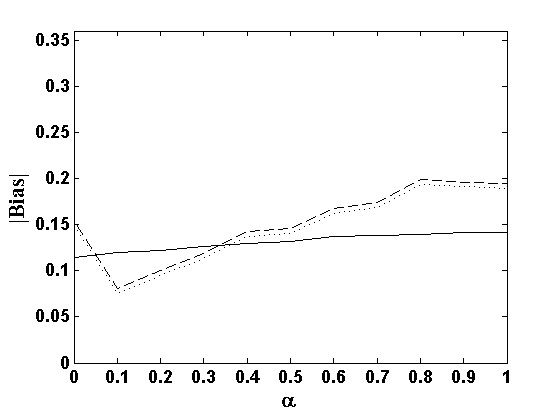}
\includegraphics[width=0.24\textwidth]{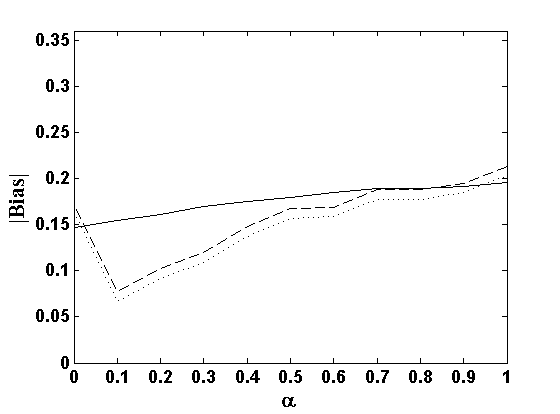}
\includegraphics[width=0.24\textwidth]{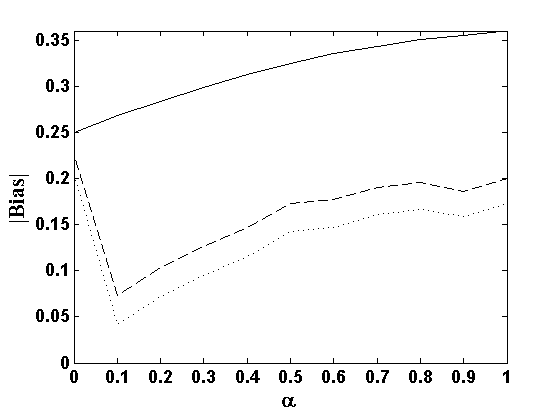}
\\
\subfloat[$k=50$]{
\includegraphics[width=0.24\textwidth]{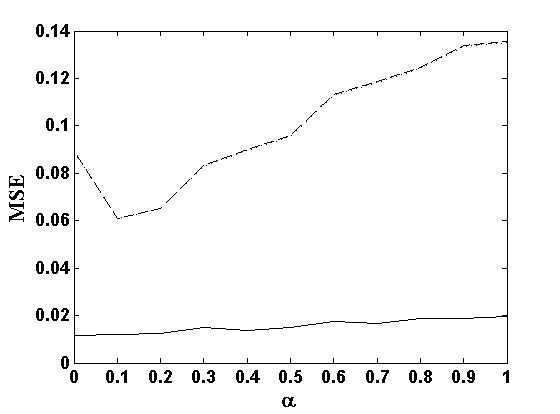}
\label{FIG:M1_5_k50}}
\subfloat[$k=150$]{
\includegraphics[width=0.24\textwidth]{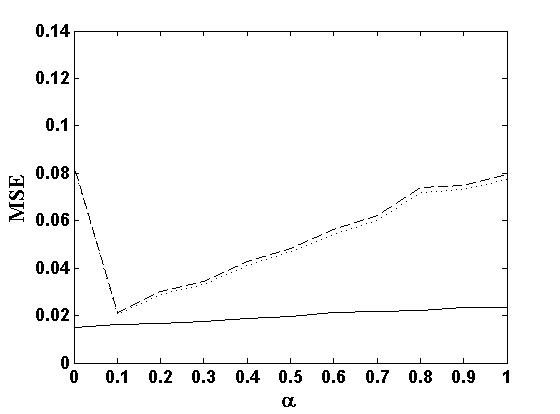}
\label{FIG:M1_5_k150}}
\subfloat[$k=250$]{
\includegraphics[width=0.24\textwidth]{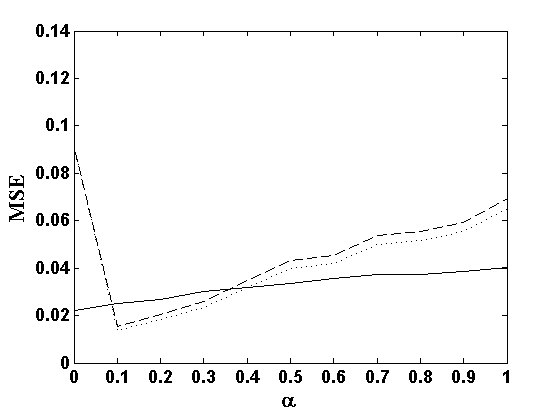}
\label{FIG:M1_5_k250}}
\subfloat[$k=500$]{
\includegraphics[width=0.24\textwidth]{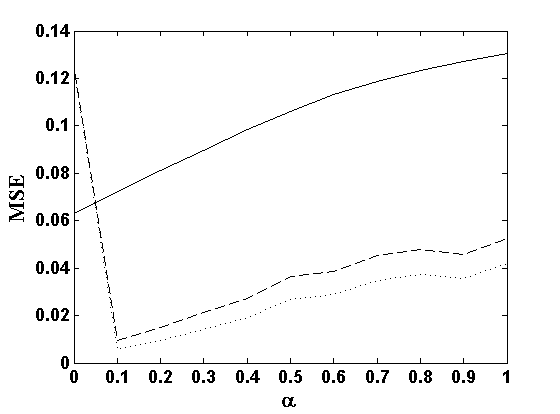}
\label{FIG:M1_5_k500}}
\caption{Empirical Bias and MSE of the estimators of $\eta$ for Model (M1') with $\epsilon=5\%$ and $n=1000$. 
[solid line: $\hat{\eta}_D^\alpha$, dotted line: $\hat{\eta}_{ER,F}^\alpha$, dashed line: $\hat{\eta}_{ER,P}^\alpha$]}
\label{FIG:M1_5}
\end{figure}

\begin{figure}[!th]
\centering
\includegraphics[width=0.24\textwidth]{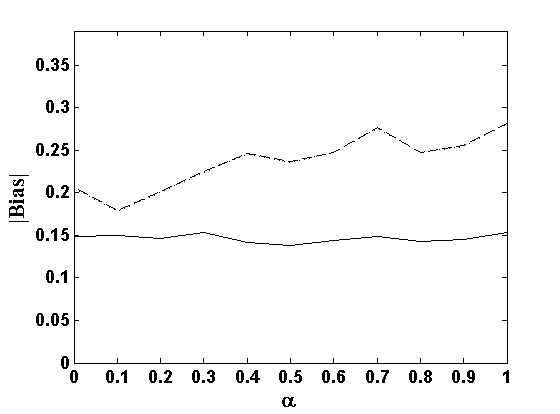}
\includegraphics[width=0.24\textwidth]{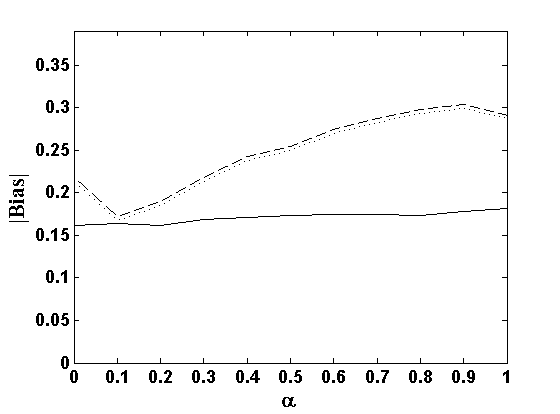}
\includegraphics[width=0.24\textwidth]{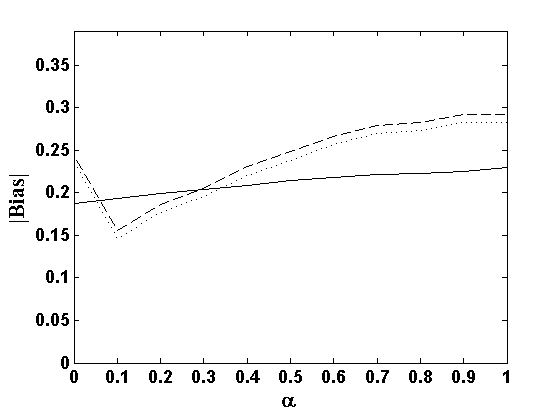}
\includegraphics[width=0.24\textwidth]{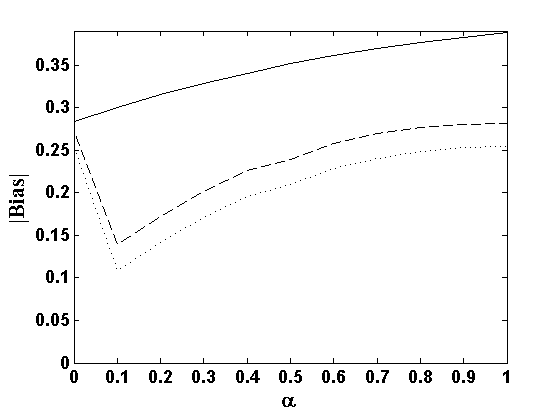}
\\
\subfloat[$k=50$]{
\includegraphics[width=0.24\textwidth]{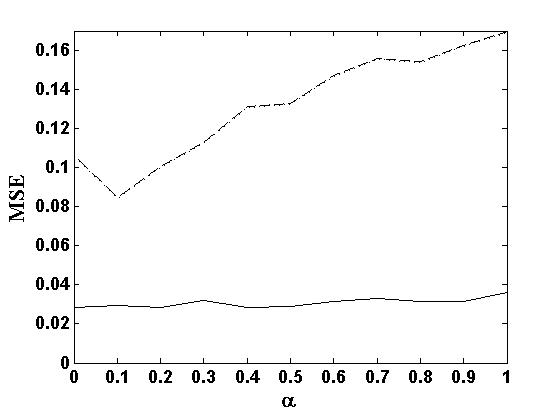}
\label{FIG:M1_15_k50}}
\subfloat[$k=150$]{
\includegraphics[width=0.24\textwidth]{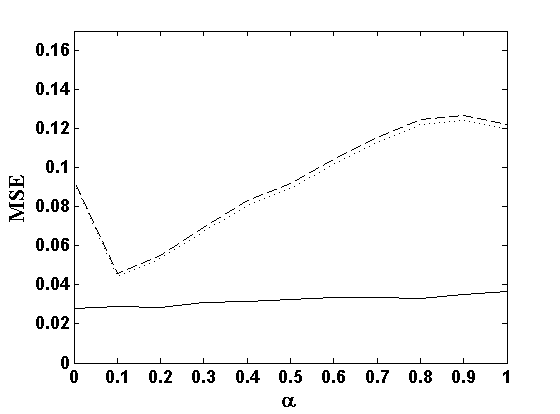}
\label{FIG:M1_15_k150}}
\subfloat[$k=250$]{
\includegraphics[width=0.24\textwidth]{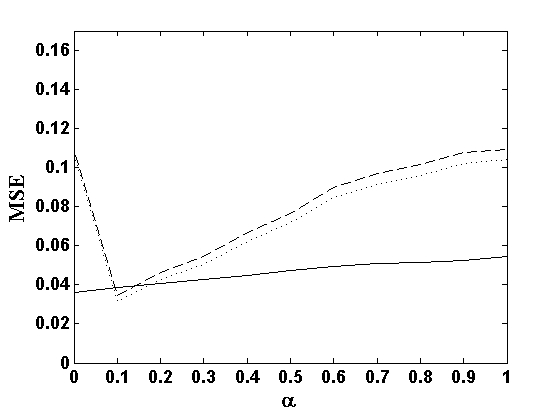}
\label{FIG:M1_15_k250}}
\subfloat[$k=500$]{
\includegraphics[width=0.24\textwidth]{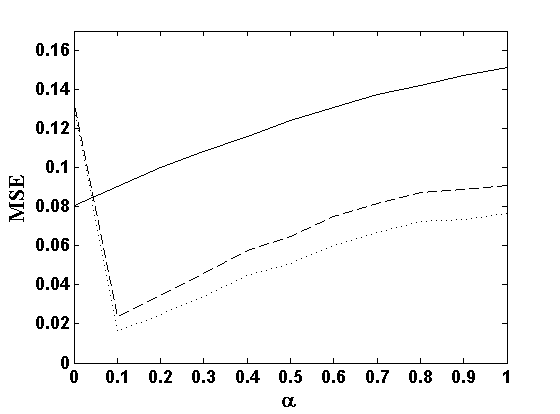}
\label{FIG:M1_15_k500}}
\caption{Empirical Bias and MSE of the estimators of $\eta$ for Model (M1') with $\epsilon=15\%$ and $n=1000$. 
[solid line: $\hat{\eta}_D^\alpha$, dotted line: $\hat{\eta}_{ER,F}^\alpha$, dashed line: $\hat{\eta}_{ER,P}^\alpha$]}
\label{FIG:M1_15}
\end{figure}

\begin{enumerate}
\item[(M1')] Contaminate (M1) by the same bivariate normal distribution but with correlation coefficient $0.75$. 
Thus, this makes the non-robust estimator of $\eta$ to be shifted upward.

\item[(M2')] Contaminate (M2) by standard bivariate normal distribution with correlation $-0.9$. 
The value of $\eta$ for contamination distribution is close to zero 
pulling the non-robust estimators downward.
\item[(M3')] Contaminate (M3) by the distribution corresponding to the Gumbel Copula with $\delta=20$ 
having tail index very close to 1.
\item[(M4')] Contaminate (M3) by the distribution corresponding to the Clayton Copula with $\delta=200$ 
(its tail index is also very close to 1).
\end{enumerate}

\begin{figure}[!th]
\centering
\includegraphics[width=0.24\textwidth]{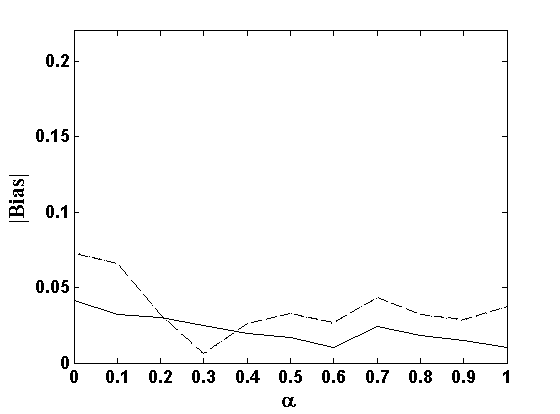}
\includegraphics[width=0.24\textwidth]{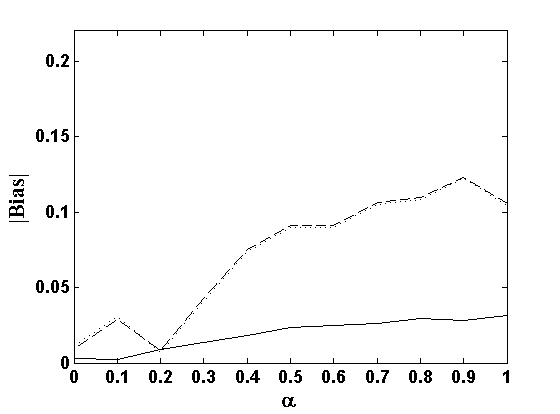}
\includegraphics[width=0.24\textwidth]{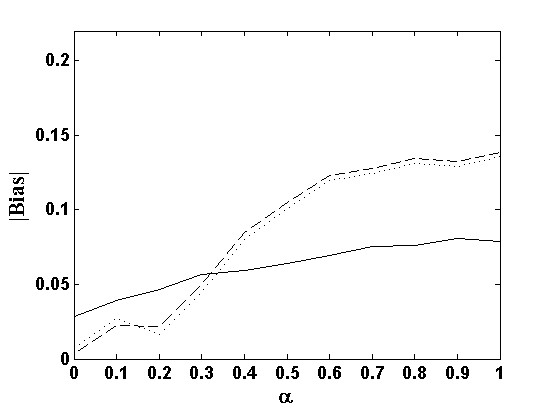}
\includegraphics[width=0.24\textwidth]{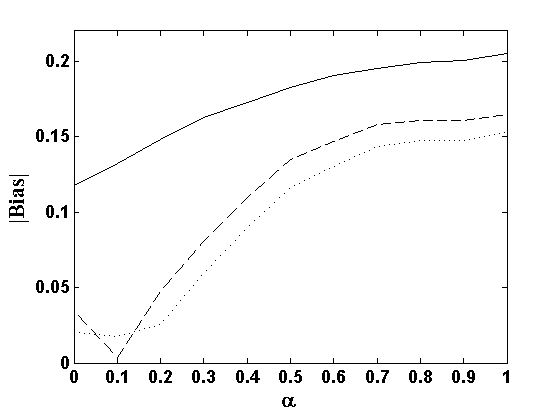}
\\
\subfloat[$k=50$]{
\includegraphics[width=0.24\textwidth]{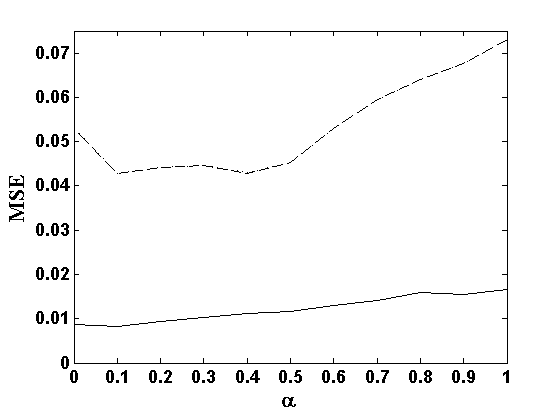}
\label{FIG:M2_5_k50}}
\subfloat[$k=150$]{
\includegraphics[width=0.24\textwidth]{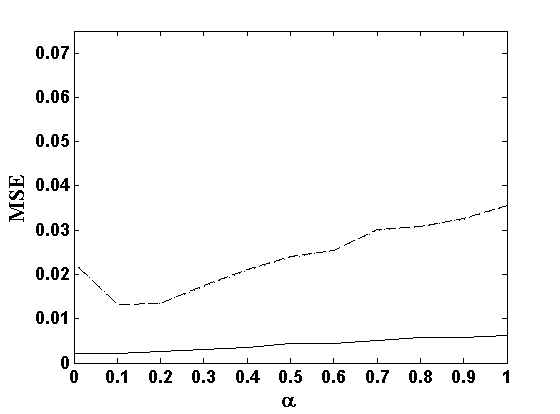}
\label{FIG:M2_5_k150}}
\subfloat[$k=250$]{
\includegraphics[width=0.24\textwidth]{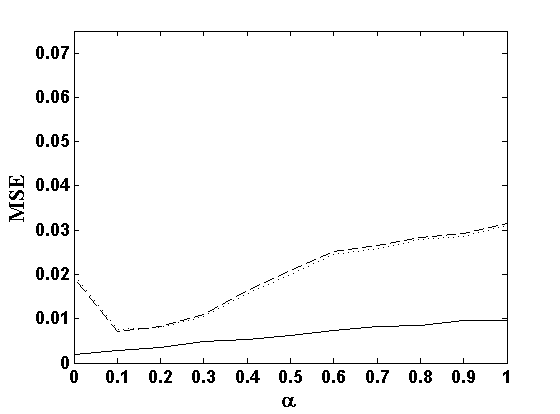}
\label{FIG:M2_5_k250}}
\subfloat[$k=500$]{
\includegraphics[width=0.24\textwidth]{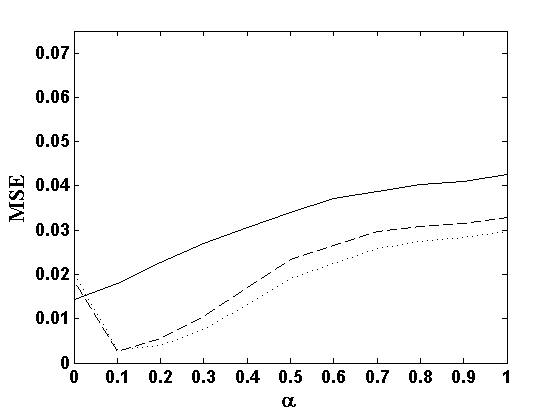}
\label{FIG:M2_5_k500}}
\caption{Empirical Bias and MSE of the estimators of $\eta$ for Model (M2') with $\epsilon=5\%$ and $n=1000$. 
[solid line: $\hat{\eta}_D^\alpha$, dotted line: $\hat{\eta}_{ER,F}^\alpha$, dashed line: $\hat{\eta}_{ER,P}^\alpha$]}
\label{FIG:M2_5}
\end{figure}

\begin{figure}[!th]
\centering
\includegraphics[width=0.24\textwidth]{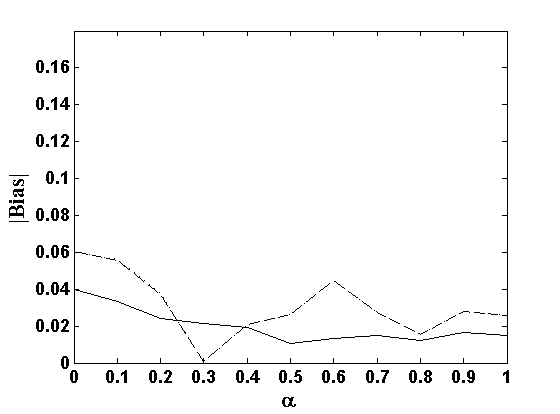}
\includegraphics[width=0.24\textwidth]{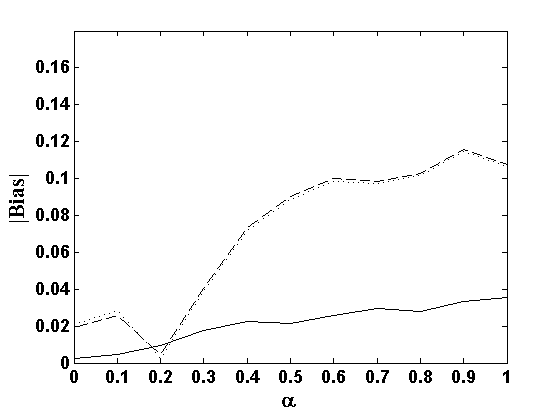}
\includegraphics[width=0.24\textwidth]{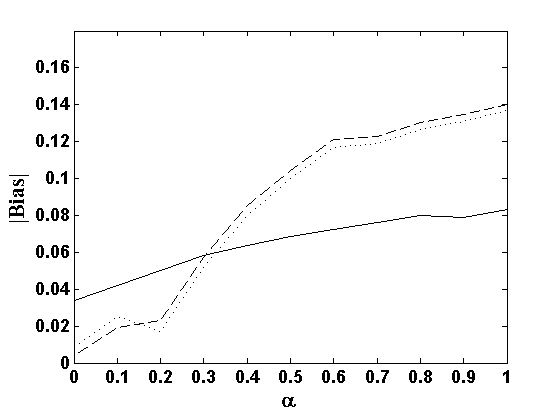}
\includegraphics[width=0.24\textwidth]{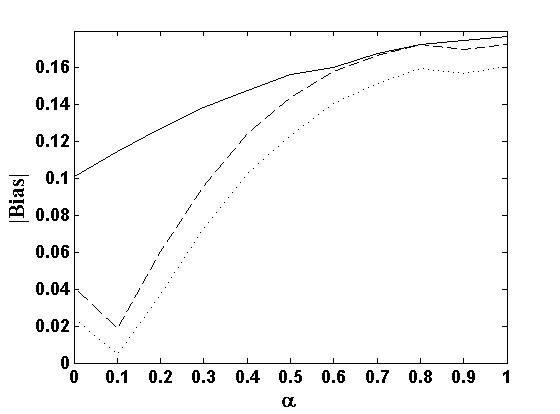}
\\
\subfloat[$k=50$]{
\includegraphics[width=0.24\textwidth]{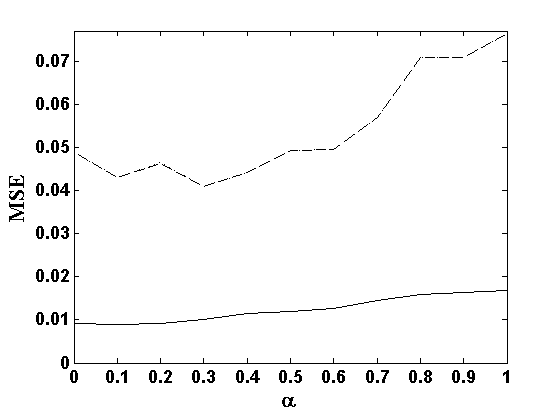}
\label{FIG:M2_15_k50}}
\subfloat[$k=150$]{
\includegraphics[width=0.24\textwidth]{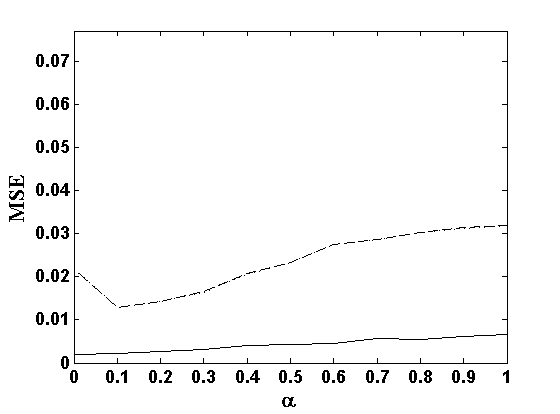}
\label{FIG:M2_15_k150}}
\subfloat[$k=250$]{
\includegraphics[width=0.24\textwidth]{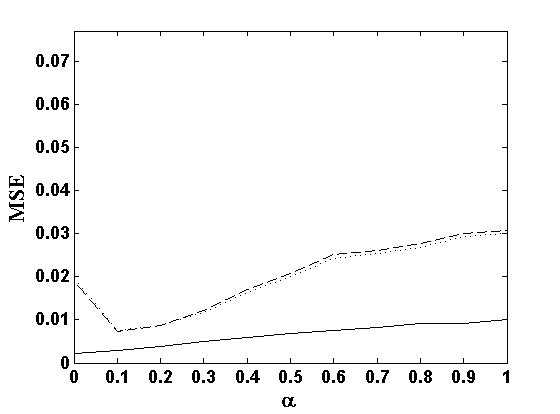}
\label{FIG:M2_15_k250}}
\subfloat[$k=500$]{
\includegraphics[width=0.24\textwidth]{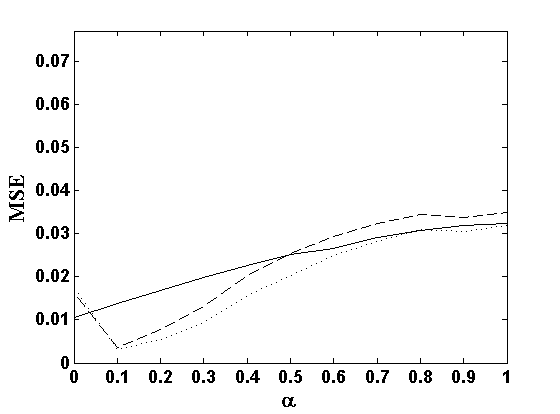}
\label{FIG:M2_15_k500}}
\caption{Empirical Bias and MSE of the estimators of $\eta$ for Model (M2') with $\epsilon=15\%$ and $n=1000$. 
[solid line: $\hat{\eta}_D^\alpha$, dotted line: $\hat{\eta}_{ER,F}^\alpha$, dashed line: $\hat{\eta}_{ER,P}^\alpha$]}
\label{FIG:M2_15}
\end{figure}

\begin{figure}[!th]
\centering
\includegraphics[width=0.24\textwidth]{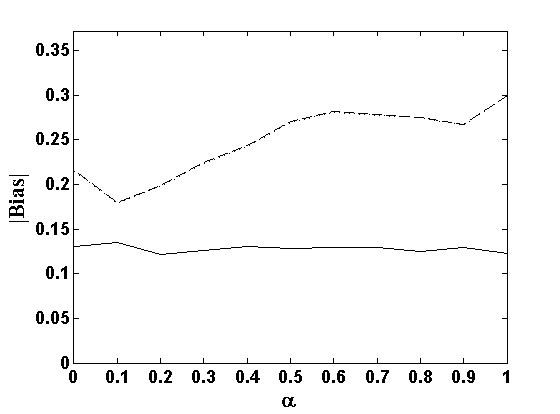}
\includegraphics[width=0.24\textwidth]{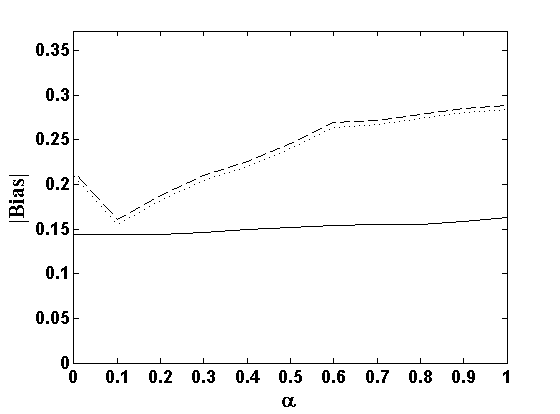}
\includegraphics[width=0.24\textwidth]{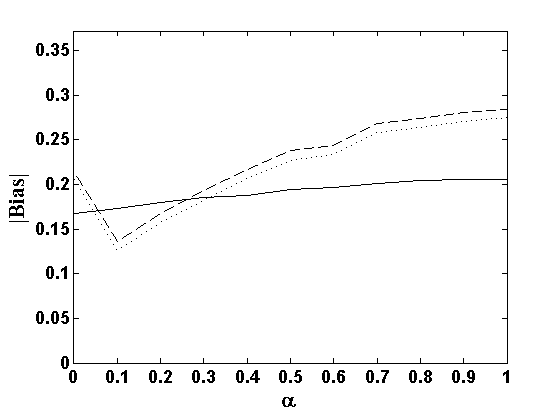}
\includegraphics[width=0.24\textwidth]{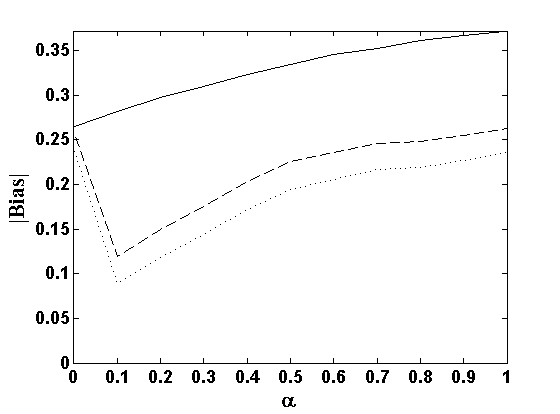}
\\
\subfloat[$k=50$]{
\includegraphics[width=0.24\textwidth]{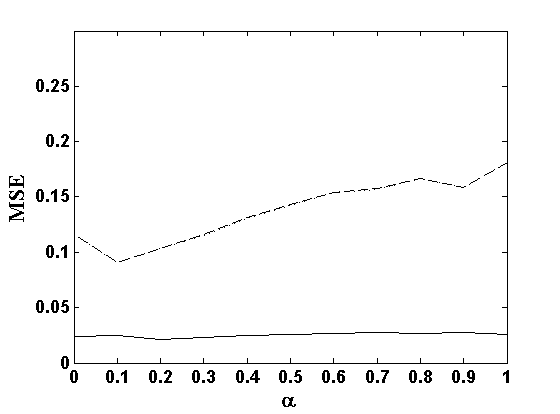}
\label{FIG:M3_5_k50}}
\subfloat[$k=150$]{
\includegraphics[width=0.24\textwidth]{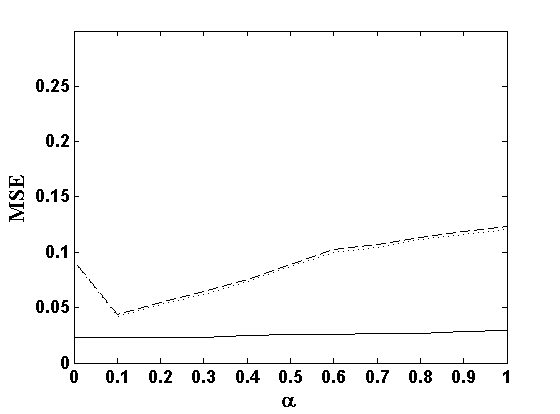}
\label{FIG:M3_5_k150}}
\subfloat[$k=250$]{
\includegraphics[width=0.24\textwidth]{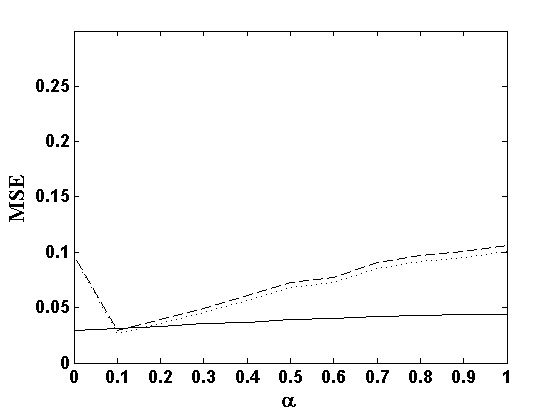}
\label{FIG:M3_5_k250}}
\subfloat[$k=500$]{
\includegraphics[width=0.24\textwidth]{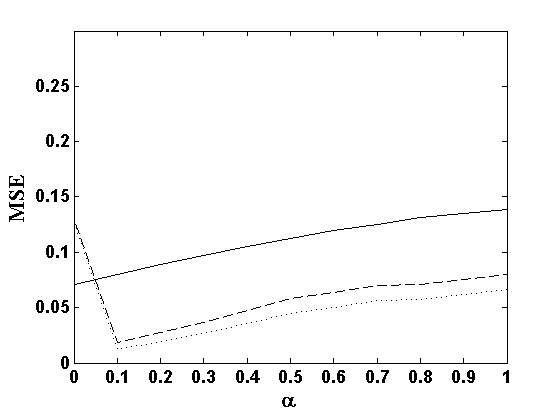}
\label{FIG:M3_5_k500}}
\caption{Empirical Bias and MSE of the estimators of $\eta$ for Model (M3') with $\epsilon=5\%$ and $n=1000$. 
[solid line: $\hat{\eta}_D^\alpha$, dotted line: $\hat{\eta}_{ER,F}^\alpha$, dashed line: $\hat{\eta}_{ER,P}^\alpha$]}
\label{FIG:M3_5}
\end{figure}

\begin{figure}[!th]
\centering
\includegraphics[width=0.24\textwidth]{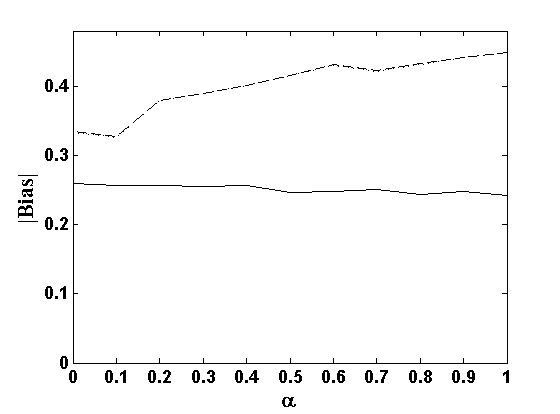}
\includegraphics[width=0.24\textwidth]{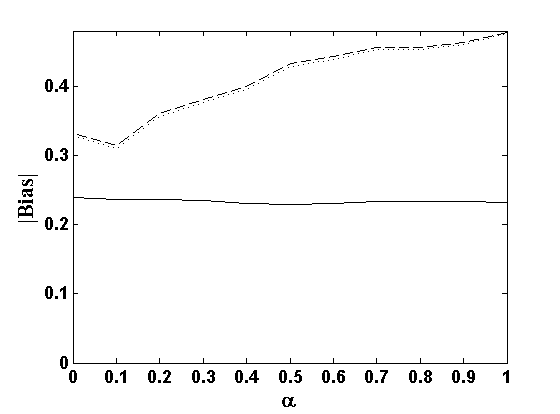}
\includegraphics[width=0.24\textwidth]{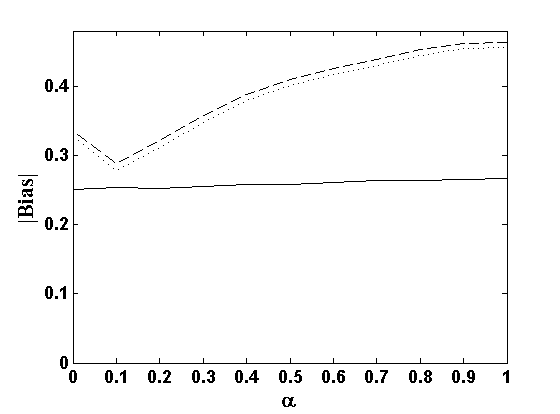}
\includegraphics[width=0.24\textwidth]{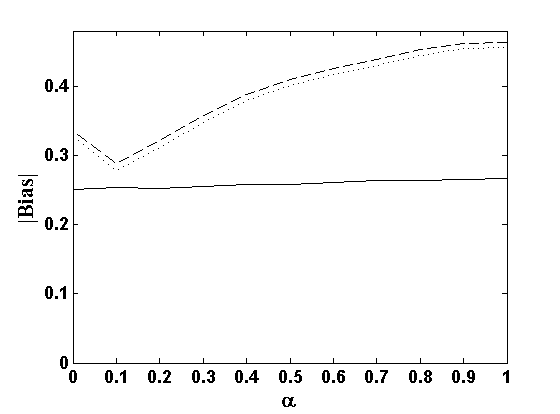}
\\
\subfloat[$k=50$]{
\includegraphics[width=0.24\textwidth]{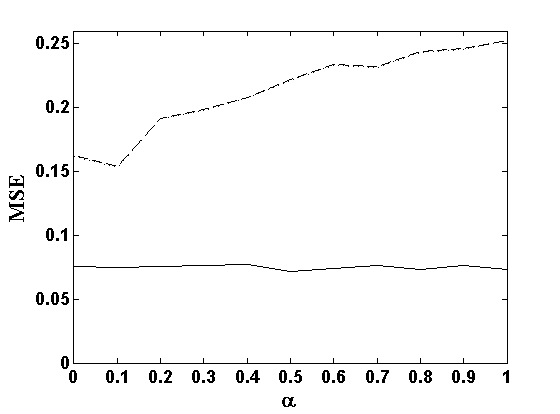}
\label{FIG:M3_15_k50}}
\subfloat[$k=150$]{
\includegraphics[width=0.24\textwidth]{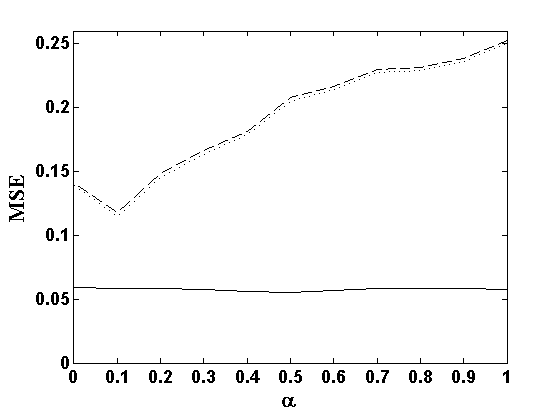}
\label{FIG:M3_15_k150}}
\subfloat[$k=250$]{
\includegraphics[width=0.24\textwidth]{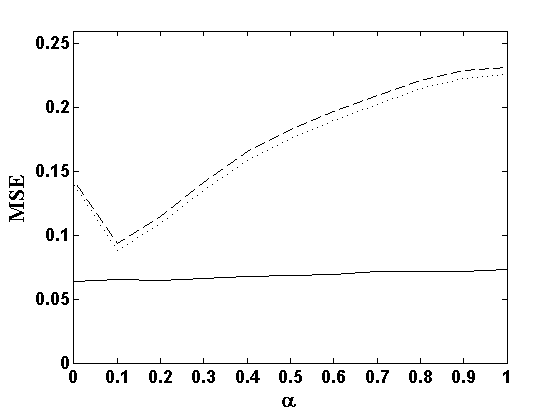}
\label{FIG:M3_15_k250}}
\subfloat[$k=500$]{
\includegraphics[width=0.24\textwidth]{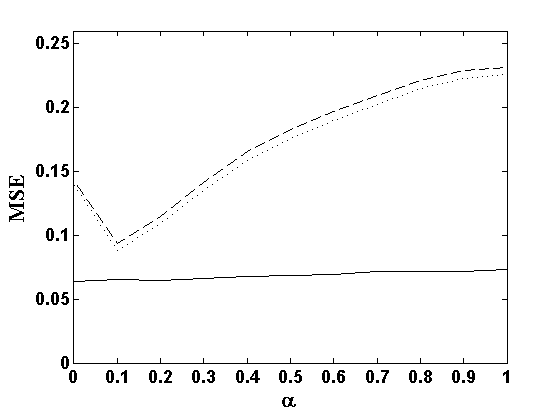}
\label{FIG:M3_15_k500}}
\caption{Empirical Bias and MSE of the estimators of $\eta$ for Model (M3') with $\epsilon=15\%$ and $n=1000$. 
[solid line: $\hat{\eta}_D^\alpha$, dotted line: $\hat{\eta}_{ER,F}^\alpha$, dashed line: $\hat{\eta}_{ER,P}^\alpha$]}
\label{FIG:M3_15}
\end{figure}

\begin{figure}[!h]
\centering
\includegraphics[width=0.24\textwidth]{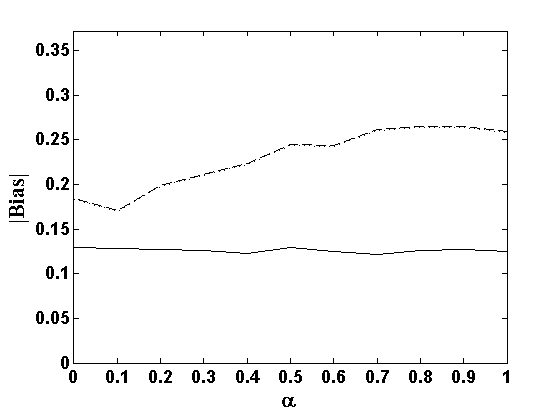}
\includegraphics[width=0.24\textwidth]{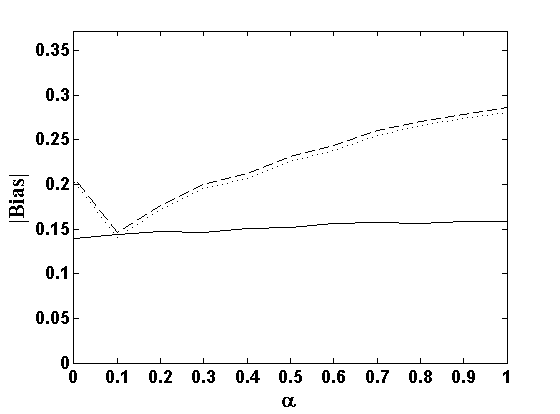}
\includegraphics[width=0.24\textwidth]{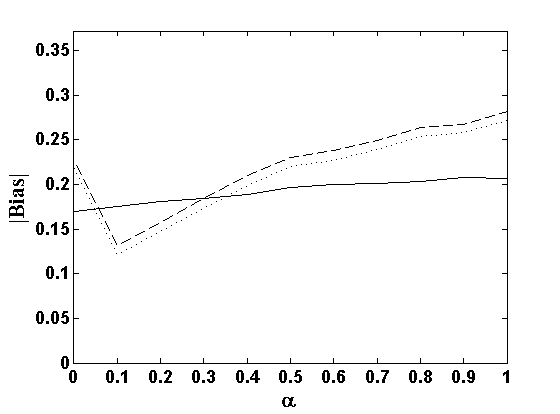}
\includegraphics[width=0.24\textwidth]{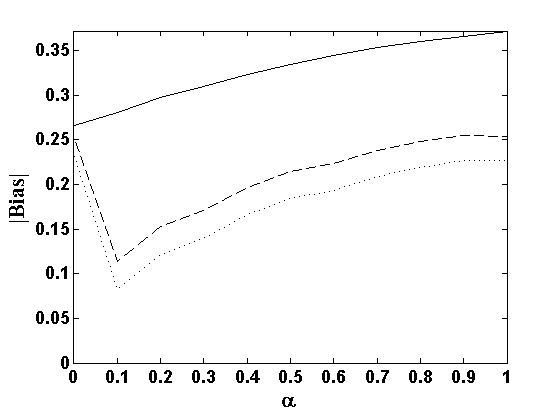}
\\
\subfloat[$k=50$]{
\includegraphics[width=0.24\textwidth]{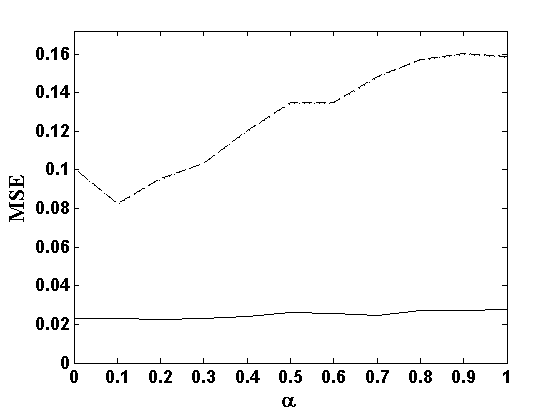}
\label{FIG:M4_5_k50}}
\subfloat[$k=150$]{
\includegraphics[width=0.24\textwidth]{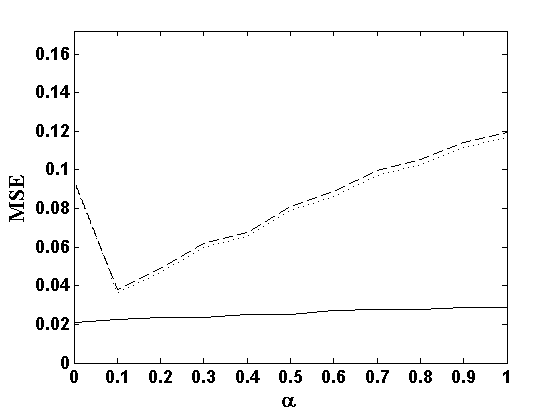}
\label{FIG:M4_5_k150}}
\subfloat[$k=250$]{
\includegraphics[width=0.24\textwidth]{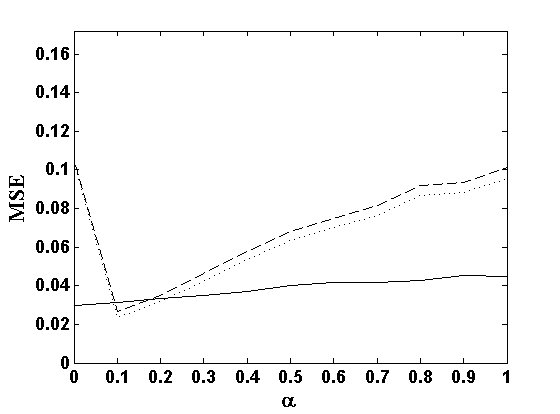}
\label{FIG:M4_5_k250}}
\subfloat[$k=500$]{
\includegraphics[width=0.24\textwidth]{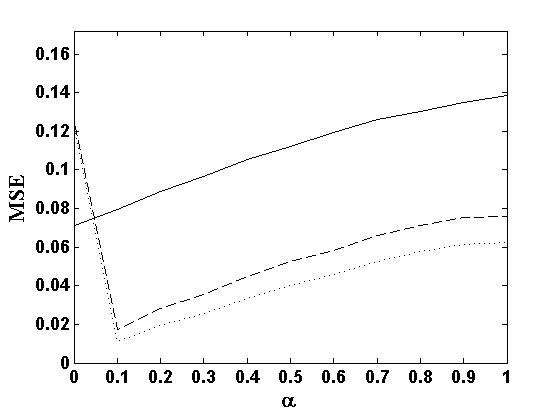}
\label{FIG:M4_5_k500}}
\caption{Empirical Bias and MSE of the estimators of $\eta$ for Model (M4') with $\epsilon=5\%$ and $n=1000$. 
[solid line: $\hat{\eta}_D^\alpha$, dotted line: $\hat{\eta}_{ER,F}^\alpha$, dashed line: $\hat{\eta}_{ER,P}^\alpha$]}
\label{FIG:M4_5}
\end{figure}

\begin{figure}[!th]
\centering
\includegraphics[width=0.24\textwidth]{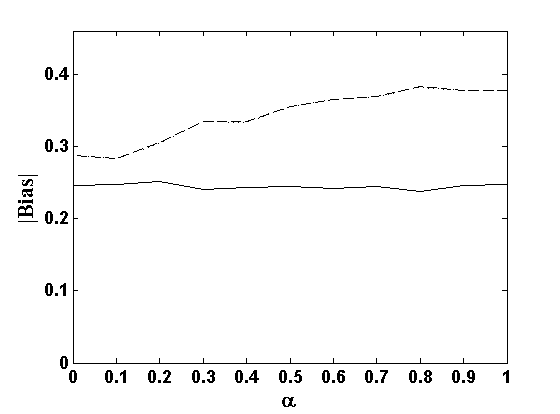}
\includegraphics[width=0.24\textwidth]{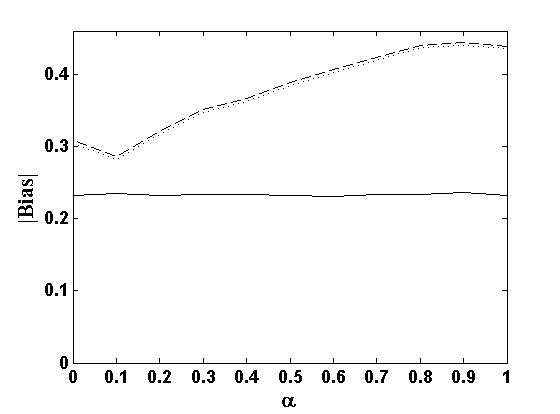}
\includegraphics[width=0.24\textwidth]{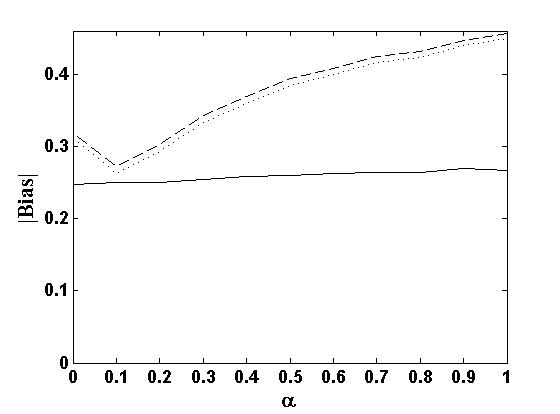}
\includegraphics[width=0.24\textwidth]{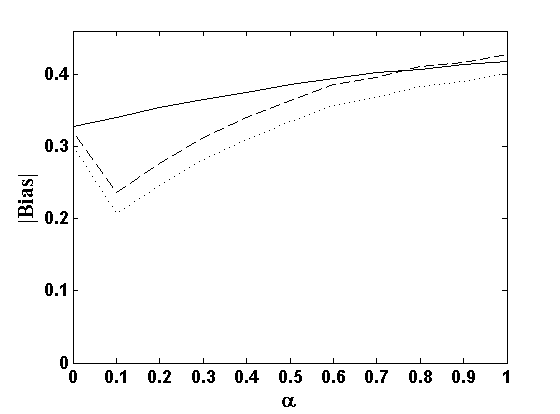}
\\
\subfloat[$k=50$]{
\includegraphics[width=0.24\textwidth]{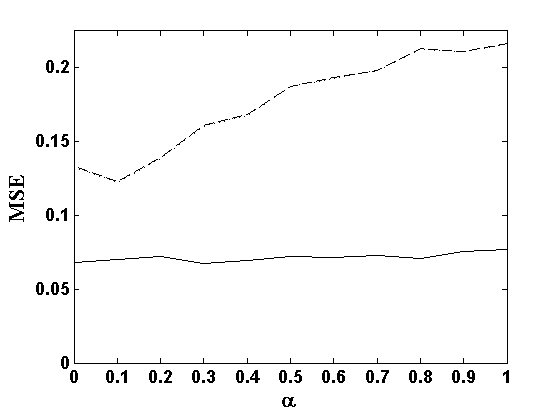}
\label{FIG:M4_15_k50}}
\subfloat[$k=150$]{
\includegraphics[width=0.24\textwidth]{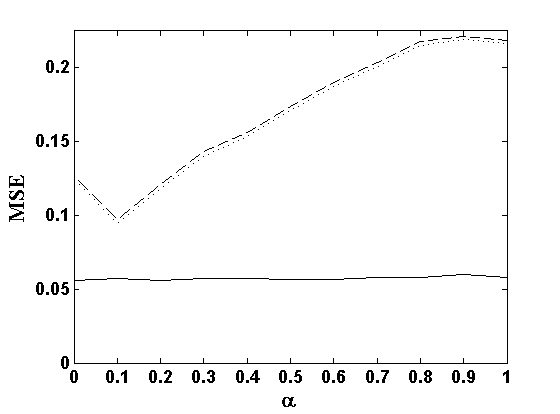}
\label{FIG:M4_15_k150}}
\subfloat[$k=250$]{
\includegraphics[width=0.24\textwidth]{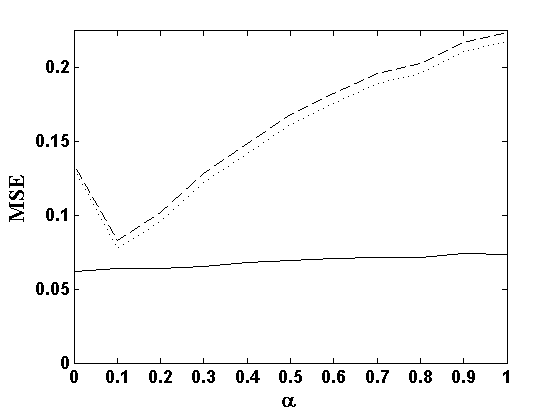}
\label{FIG:M4_15_k250}}
\subfloat[$k=500$]{
\includegraphics[width=0.24\textwidth]{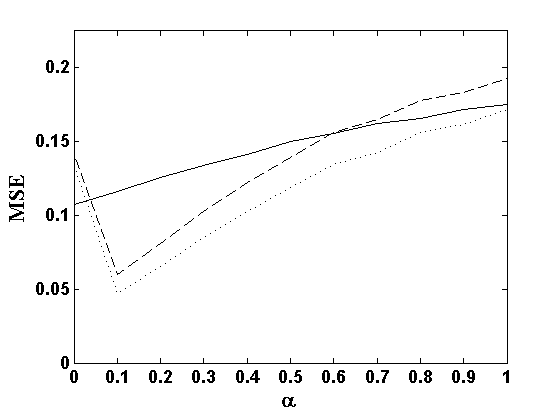}
\label{FIG:M4_15_k500}}
\caption{Empirical Bias and MSE of the estimators of $\eta$ for Model (M4') with $\epsilon=15\%$ and $n=1000$. 
[solid line: $\hat{\eta}_D^\alpha$, dotted line: $\hat{\eta}_{ER,F}^\alpha$, dashed line: $\hat{\eta}_{ER,P}^\alpha$]}
\label{FIG:M4_15}
\end{figure}

For these four contaminated models, we again compute the empirical bias and MSE based on $1000$ replications taking 
the contamination proportion $\epsilon$ to be $5\%$ (small contamination) as well as $15\%$ (heavy contamination).
Again we only present the results for $n=1000$ in Figures \ref{FIG:M1_5} to \ref{FIG:M4_15} respectively. 
Interestingly, even under contamination, the performance of the estimators $\hat{\eta}_{ER,P}^\alpha$ and 
$\hat{\eta}_{ER,F}^\alpha$ are quite similar in most of the cases under consideration.

For the contaminated model (M1') with $5\%$ and $15\%$ contaminations, 
the performance of the estimators $\hat{\eta}_{ER,F}^\alpha$  with respect to the values of 
$\alpha$ and $k$ are quite similar to the unrestricted cases. However, the extend of bias and MSE increases 
as the contamination proportion increases and hence the optimum choice of tuning parameter $k$ becomes larger ---
the extend of improvement over existing ones at the unrestricted optimum choice $k=250$ and $\alpha=0.1$ 
decreases with increasing contamination. On the other hand, bias of the estimator $\hat{\eta}_{D}^\alpha$ 
increases with $\alpha$ slightly for smaller $k$ at $5\%$  contamination but their MSE remains 
almost the same over $\alpha$; in case of $15\%$ contamination, both bias and MSE of 
$\hat{\eta}_{D}^\alpha$ decreases slightly with increasing $\alpha$ at smaller values of $k$. 
For the model (M2') with $5\%$ and $15\%$ contamination again, the nature of the estimators 
$\hat{\eta}_{ER,F}^\alpha$  and $\hat{\eta}_{D}^\alpha$ are quite similar to the unrestricted case (M2). 
For the contaminated models (M3') and (M4') with $5\%$ contamination, 
where the true values of $\eta$ are same to (M1') but the contamination structures are different, 
the performance of all the estimators are similar to that under (M1') except that 
the extend of their bias and MSE are more compared to that under (M1'). 
At $15\%$ contamination, this difference in the extend of bias and MSE further increases, 
keeping the overall pattern same; but only for model (M3'), the estimator $\hat{\eta}_{ER,F}^\alpha$ 
can not perform better than  $\hat{\eta}_{D}^\alpha$ even at $k=500$ although 
they gives better results than the existing  Matthyas and Beirlant (2001) estimator at any $k$ and $\alpha=0.1$.

Combining all the findings on simulation study, 
it is quite clear that $\hat{\eta}_{D}^\alpha$ always performs better for smaller values of $k$, 
whereas $\hat{\eta}_{ER,F}^\alpha$  and $\hat{\eta}_{ER,P}^\alpha$ 
perform similar to each other with lower MSE at larger values of $k$; 
this nature of the estimators also remains stable under the contamination in data 
and robustness increases as $\alpha$ increases. 
In fact, the $\hat{\eta}_{ER,F}^\alpha$  or $\hat{\eta}_{ER,P}^\alpha$ at $\alpha$ near $0.1$ and 
some optimum $k$ perform much better with respect to both bias and MSE than the other estimators 
proposed here along with the existing estimators like Hill's estimator and Matthyas and Beirlant (2001) estimator 
under presence of contamination as well as under pure data. In fact large values of $k$ leads to less
variation while smaller values produce less bias; we have also observed from the influence function analysis
that the bias-robustness of the proposed estimators increases with increasing values of $k$.
The optimum choice of $k$ based on a trade-off between bias and variance turns out to be near $k = 250 = n/4$ 
for most cases considered in this paper and sometime slightly more (but $\le 500 = n/2$).
However, it is to be noted that the proposed estimators are constructed under two sets of assumptions, 
which need to be taken care of before applying the corresponding estimators.

\section{Conclusion}
This work considers the problem of estimating the bivariate tail dependence coefficient and 
proposes some robust estimators under suitable assumptions and model approximation borrowing 
the idea from the univariate tail index estimation.
One set of proposed estimators generalizes the classical Hill's estimator 
and another generalizes the popular Matthyas and Beirlant (2001) estimator having lesser bias 
to generate their robust version. 
The estimators use the special structure of density power divergence to achieve robustness 
which is then measured through the classical influence function analysis.
It is illustrated through an extensive  simulation study that the proposed estimators perform 
much better than the existing estimators under contamination in data; 
even under the pure data some member of them outperform the existing estimators and have lesser biases.
Further, noting that the performances of the proposed estimators depends on the tuning parameter $k$ and $\alpha$,
we have presented some empirical guidance on their choices based on the simulation study.
This enriches the proposed robust estimation procedure in its practical applicability on any real data.  
Therefore, the present paper proposes robust estimators of the bivariate tail dependence index 
that can be applied for solving any practical problem with outliers in data with 
some ``good" properties as illustrated here and opens a new door in the context of robust 
multivariate extreme value analysis for future research.

\end{document}